\documentclass[pra,twocolumn,superscriptaddress,longbibliography]{revtex4-1}
\usepackage{amsmath,amssymb}
\usepackage{mathtools}
\usepackage{simplewick}

\usepackage[applemac]{inputenc}
\usepackage[english]{babel}
\usepackage[pdftex]{graphicx}
\usepackage{color}
\usepackage{dcolumn}
\usepackage{bm}
\usepackage{mathbbol}
\usepackage{array,multirow,graphicx}
\usepackage{amsfonts}
\usepackage{times}
\usepackage{xcolor}

\usepackage{hyperref}
\usepackage{cleveref}

\newcommand{\ket}[1]{\left\vert #1\,\right\rangle}
\newcommand{\bra}[1]{\left\langle #1\,\right\vert}

\begin{document}

\title{Timescales in the quench dynamics of many-body quantum systems: \\ Participation ratio vs out-of-time ordered correlator}
\author{Fausto Borgonovi}
\affiliation{Dipartimento di Matematica e
  Fisica and Interdisciplinary Laboratories for Advanced Materials Physics,
  Universit\`a Cattolica, via Musei 41, 25121 Brescia, Italy}
\affiliation{Istituto Nazionale di Fisica Nucleare,  Sezione di Pavia,
  via Bassi 6, I-27100,  Pavia, Italy}
\author{Felix M. Izrailev}
\affiliation{Instituto de F\'{i}sica, Benem\'{e}rita Universidad Aut\'{o}noma
  de Puebla, Apartado Postal J-48, Puebla 72570, Mexico}
\affiliation{Department of Physics and Astronomy, Michigan State University, E. Lansing, Michigan 48824-1321, USA}
\author{Lea F. Santos}
\affiliation{Department of Physics, Yeshiva University, New York,
New York 10016, USA}

\date{\today}

\begin{abstract}

We study quench dynamics in the many-body Hilbert space using two isolated systems with a finite number of interacting particles: a paradigmatic model of randomly interacting bosons and a dynamical (clean) model of interacting spins-$1/2$. For both systems in the region of strong quantum chaos, the number of components of the evolving wave function, defined through the number of principal components $N_{pc}$ (or participation ratio), was recently found to increase exponentially fast in time [Phys. Rev. E {\bf 99}, 010101R (2019)]. Here, we ask whether the out-of-time ordered correlator (OTOC), which is nowadays widely used to quantify instability in quantum systems, can manifest analogous time-dependence. We show that $N_{pc}$ can be formally expressed as the inverse of the sum of all OTOC's for projection operators.  While none of the individual projection-OTOC's shows an exponential behavior, their sum decreases exponentially fast in time. The comparison between the behavior of the OTOC  with that of the $N_{pc}$ helps us better understand wave packet dynamics in the many-body Hilbert space, in close connection with the problems of thermalization and information scrambling.
\end{abstract}

\maketitle

\section{Introduction}

There is currently great interest in the study of non-equilibrium quantum dynamics of isolated systems with many interacting particles. This is partially justified by significant experimental progress that makes possible the study of the coherent evolution of many-body quantum systems for long times~\cite{Kaufman2016,LewisARXIV,WeiARXIV}. Yet, despite important analytical and experimental advances, several questions remain open. A timely discussion refers to the conditions~\cite{Borgonovi2016,Alessio2016} and timescales~\cite{Borgonovi2019,SchiulazARXIV,DymarskyARXIVThouless} for the onset of equilibration and thermalization that can emerge without the influence of an environment. When studying these topics, one should distinguish systems at the thermodynamic limit, addressed by mean-field theories~\cite{Eriksson2018}, from systems with a finite number of particles. The latter situation emerges in experiments with cold atoms and ion traps, where the number of particles can be small and controlled. 

Analytical breakthroughs in the study of many-body quantum dynamics have been recently achieved in high energy physics~\cite{Maldacena2017FP}, where quantum systems without gravity are equated to classical gravitational systems in a higher spatial dimension.  A quantity that became central in many of these studies is the out-of-time-order correlator (OTOC), first introduced in the semiclassical analysis of superconductivity in Ref.~\cite{Larkin1969}.  Existing analytical results for the evolution of the OTOC have been obtained by taking the average in the canonical ensemble~\cite{Sekino2008,KitaevTALK,Maldacena2016PRD,Maldacena2016JHEP}, thus assuming implicitly the thermodynamic limit. The present work focuses on the dynamics of finite isolated systems with interacting Bose or Fermi particles and employs the OTOC to describe the gradual spreading of the initial wave packet in the many-body Hilbert space.

The OTOC can be measured experimentally with nuclear magnetic resonance platforms and ion traps~\cite{Garttner2017,Li2017,NiknamARXIV}. Among various applications, it has been used to quantify the spread of quantum information~\cite{Swingle2018} and the exponential instability of quantum systems that have a chaotic classical counterpart, as supported by semiclassical analysis~\cite{Rammensee2018,Jalabert2018}. This has given birth to another method to detect chaos in quantum dynamics, a goal pursued by several earlier works~\cite{Peres1996,Levstein1998,Cucchietti2002,Gorin2006,Elsayed2015}.

The quantum-classical correspondence between the exponential growth rate of the OTOC and the classical Lyapunov exponent has being numerically corroborated for finite systems with few degrees of freedom, such as one-body chaotic systems~\cite{Rozenbaum2017,RozenbaumARXIV} and the Dicke model with two degrees of freedom~\cite{Chavez2019}. However, little is known about this correspondence for finite  quantum systems with many interacting particles. Studies of the OTOC have contributed to a significant renewed interest in the problem of the quantum-classical correspondence for chaotic systems, which is a study initiated about 40 years ago with the investigation of one-body chaos. 
 
In the paradigmatic Kicked Rotator (KR) model, it was found numerically \cite{Casati1979} and explained analytically \cite{Chirikov1981,Shepelyansky1983} that there are {\it two timescales} on which one can speak of the quantum-classical correspondence for the dynamics of wave packets. One is the timescale due to the Ehrenfest theorem according to which the center of the wave packet in phase space follows, for some time, the corresponding classical trajectories. In the case of strong chaos, the timescale $t_E$ for this correspondence  was analytically studied in Refs.~\cite{Berman1978,Zaslavsky1981} and shown to be proportional to $\ln (1/\hbar)$, where $\hbar$ stands for an effective dimensionless Planck constant. The other timescale, $t_D$, is due to the {\it dynamical localization} occurring in the momentum space of the KR \cite{Casati1979,Chirikov1981,Chirikov1988,Izrailev1990}. The second moment of the wave packet in momentum space nicely mimics classical diffusion on the timescale $t_D \propto 1/\hbar^2$, which is much longer than $t_E$. It was later argued that this localization may be compared with the Anderson localization in 1D disordered models with long-range hopping~\cite{Fishman1982} and the localization in quasi-1D random models described by band random matrices \cite{Casati1990,Casati1991,Zyczkowski1992,Izrailev1995,Izrailev1996}.  

The importance of these old results obtained for the KR is two-fold. First, they show that the classical diffusion coefficient is related to the localization length of the quasienergy eigenfunctions in momentum space~\cite{Chirikov1981,Shepelyansky1986}, which is a pure quantum concept. Second, they demonstrate that the timescale for the quantum-classical correspondence can be very different for different observables. As mentioned above, global observables, such as the second moment of the probability distribution in momentum space, can coincide with their classical counterparts on a timescale much larger than that defined by the Ehrenfest theorem. This point is of special relevance for studies of the evolution of observables in many-body systems. A question of  particular interest is how the number $N$ of quantum particles enters the characteristic timescales involved in the scrambling of information, equilibration, and thermalization~\cite{Borgonovi2019,SchiulazARXIV,DymarskyARXIVThouless}. 

It was shown in~\cite{Borgonovi2019} that when the eigenstates of a many-body quantum system are strongly chaotic, the number of principal components $N_{pc}$ (or participation ratio) involved in the dynamics of the wave function in the many-body Hilbert space increases exponentially fast in time.  The growth rate  was found to be $2\Gamma$, where $\Gamma$ is the energy width of the strength function. This function, introduced in nuclear physics and known in solid state physics as local density of states (LDOS), is defined by projecting an unperturbed many-body state onto the basis defined by the total Hamiltonian that includes the inter-particle interaction. Knowledge of the LDOS is very important in the analysis of quench dynamics, since its Fourier transform is the survival probability, which describes the decay of the initial state. 

The exponential growth of $N_{pc}$  lasts for some time $t_S$ before the saturation of the dynamics, which happens due to the finite size of the many-body Hilbert space. It was found in~\cite{Borgonovi2019} that, for a large number of particles, $N\gg 1$,  the saturation time is approximately given by $t_S \propto N\hbar/ \Gamma$. Since $t_S$ is proportional to the number of particles  $N$, it  can be much larger than the characteristic time for the depletion of the initial state given by  $\hbar/\Gamma$. The timescale $t_S$ represents the time for thermalization~\cite{Borgonovi2019}, according to which an initial wave packet ergodically fills the  energy shell~\cite{casati1993,casati1996,Izrailev2001}. The spread of the initial state reflects the delocalization of the energy eigenstates, which is due to the strong inter-particle interactions~\cite{Santos2012PRL,Santos2012PRE}. These states do not fill the whole Hilbert space, just the part defined by the inter-particle interaction. 

In the present work, we explore the relationship between $N_{pc}$  and a particular kind of OTOC. The former quantifies the number of unperturbed many-body states that contribute to the evolution of the wave packet, while the OTOC measures the degree of non-commutativity in time between two different Hermitian operators. In the literature, these are usually taken as local operators in real space. Here, we use instead projection operators in the many-body Hilbert space, which are local in this space. We show that the inverse of the sum of all OTOC's coincides with $N_{pc}$. 

In our analysis, we distinguish between two categories of OTOC's: the autocorrelator, where both projections are made on the initial state, and the case involving a projection onto a many-body state other than the initial state, referred to as projection-OTOC. While the autocorrelator decays exponentially as $e^{-2 \Gamma t}$, we find that a single projection-OTOC does not exhibit exponential behavior. However, when we look at the sum of all projection-OTOC's, we find a non-monotonic behavior in time, where an initial growth is followed by an exponential decay. This decay happens within the time interval of the exponential increase of $N_{pc}$.

We consider two models, the well-known two-body random ensemble (TBRE) with a finite number of bosons interacting randomly and a dynamical (deterministic) one-dimensional (1D) spin-$1/2$ model with nearest and next-nearest neighbor couplings only. The TBRE falls into the broader category of the so-called embedded ensembles, which have been thoroughly studied since the 1970's in the context of nuclear physics and quantum chaos \cite{Brody1981,Kota2001,KotaBook}. The Sachdev-Ye-Kitaev (SYK) models \cite{KitaevTALKkitp,Sachdev1993}, which have received increasing attention in high energy physics,  are also examples of embedded random ensembles. For both models that we study, we choose parameters for which the eigenstates involved in the dynamics are composed by a very large number of unperturbed many-body states. 

The paper is organized as follows. In Sec.~\ref{Sec:models}, we describe the two models considered. Section~\ref{Sec:PR-OTOC} presents the relationship between OTOC and $N_{pc}$. In Sec.~\ref{Sec:An-Num}, we show analytical as well as numerical results for both the TBRE and the spin model. In Sec.~\ref{Sec:Conclusions}, we summarize our results and discuss some possible future directions. 

\section{Models and Quench Dynamics}

\label{Sec:models}
We consider a bosonic TBRE and a 1D spin-1/2 system, both of them described by the Hamiltonian 
\begin{equation}
H=H_0+V,
\end{equation} 
where 
\[
H_0 = \sum_k E_k^0 \ket{k}\bra{k}
\]
stands for the unperturbed (integrable) part of the total Hamiltonian $H$, with
\[
H =\sum_\alpha E^\alpha \ket{\alpha}\bra{\alpha},
\]
and $V$ represents the two-body interactions.
In what follows we set $\hbar =1$. We focus on the case where the perturbation $V$ is sufficiently strong, so that a large part of the energy spectrum of $H$ contains chaotic eigenstates. 

Since our study concentrates on the dynamics occurring in the unperturbed many-body space of chaotic systems, a definition of what we mean by quantum chaos is in order. For one-body systems, it is common lore to associate quantum chaos with level statistics described by full random matrices. However, in realistic finite many-body models, not all eigenstates are random vectors, as in full random matrices, and not all of them are involved in the dynamics. Therefore, spectrum statistics obtained by taking into account {\it all} eigenvalues is not the best way to characterize the dynamics, which is only due to those eigenstates that are present in an initially excited wave packet. Our approach to quantum chaos is linked with the structure of the eigenstates. They are called chaotic when they are fully delocalized in the energy shell and are composed of many uncorrelated components (see, for example, Refs.~\cite{Santos2012PRL,Santos2012PRE}).

\subsection{Two-body random ensemble}
The TBRE describes $N$ identical bosons occupying $M$ single-particle levels; the latter are specified (and reordered) by random energies $\epsilon_s$. The mean spacing $\langle \epsilon_s- \epsilon_{s-1} \rangle \equiv \delta = 1 $ sets the energy scale defining the width of the unperturbed energy spectrum, $NM\delta$. The choice to have random single-particle energies is not a necessary condition for the results obtained below. It is used to remove the degeneracy in the unperturbed many-body spectrum.  

 The Hamiltonian of the TBRE is written as, 
\begin{equation}
  H= \sum_{s=1}^M  \epsilon_s \, a^\dag_s a_s  +
 \sum_{s_1,s_2,s_3,s_4=1}^M V_{s_1 s_2 s_3 s_4} \, a^\dag_{s_1} a^\dag_{s_2} a_{s_3} a_{s_4},
\label{ham}
\end{equation}
where $a_s$ ($a_s^{\dagger}$) is the annihilation (creation) operator on the single-particle energy level $\epsilon_s$, so the number operator $ n_s = a^\dagger_s a_s $ gives the probability for the occupation of the $s$-th single-particle energy level, $n_s/N$. The two-body matrix elements $ V_{s_1 s_2 s_3 s_4} $  are Gaussian random entries with  zero mean and variance ${\cal V}^2$. The Hamiltonian conserves the total number of bosons, so the analysis is done for a single subspace of dimension 
\[
{\cal D} = \dfrac{(N+M-1)!}{N!(M-1)!}.
\]
Throughout the paper, we fix the number of  single-particles levels, $M=11$, and we vary the number of particles $N$ from 4 to 8.
That corresponds to a size ${\cal D}$ of the many-body space  ranging from 1001 up to  43758. The strength $V$ of the inter-particle interaction is chosen so that ${\cal V} = 0.4$ to have a large energy region with strongly chaotic eigenstates~\cite{Borgonoviaip}. The eigenstates $\ket{k} $ of $H_0$ constitute the unperturbed many-body basis (also called mean-field basis) in which we study the dynamics of the wave packets and in second quantized form they can be written as $\ket{n_1,...,n_s,...,n_M}$ where $n_s$ is the number of bosons in the $s$-th single-particle energy level. 

The TBRE Hamiltonian matrix is very sparse, because only a fraction of the unperturbed many-body states of $H_0$ are directly connected  by the two-body interaction $V$. The number of non-zero off-diagonal matrix elements ${\cal N} $ depends on the particularly chosen matrix line, but it is generally much smaller than the total matrix dimension $\cal{D}$. It is not possible to give a general analytical expression for ${\cal N}$, but upper and lower bounds as a function of $N,M$ have been estimated as follows~\cite{Borgonoviaip}, 
\begin{equation}
\label{meff-max}
\frac{(M-1)(M+2)}{2}  \leq {\cal N} \leq  N(M-1) \left[ 1 + \frac{(N-1)(M-2)}{4}  \right] .
\end{equation}
In particular, the minimal number of directly coupled states, which is independent of $N$,  is obtained when all $N$ particles occupy only one single-particle energy level.
Another feature of the TBRE matrices is their band-like structure, which causes the eigenstates close to the ground state to be much less delocalized than the states closer to the center of the spectrum. 

The TBRE was originally developed to explain the statistical properties of  complex systems with interacting Fermi-particles, such as highly excited nuclei and molecules~\cite{French1970,Brody1981}. It was later applied to systems of interacting bosons, to which, in the dilute limit, many aspects of energy spectra and eigenstates are similar to those of systems of random interacting fermions. To date, it has been extensively investigated for fermions~\cite{Flambaum1997,Altshuler1997} and  for bosons~\cite{Kota2001,Kota2001PRE,KotaBook,Benet2003}. This model is a particular case of the embedded ensembles with $q$-body interactions.  When $q=2$ we have the TBRE and when $q=N$, we recover the full random matrices. 

In contrast to the standard ensembles of full random matrices, TBREs are much closer to realistic physical systems, since they take into account the two-body nature of the interactions, the type of interacting particles (fermions or bosons), the strength of the inter-particles interaction, and the  properties of single-particle spectra. 

\subsection{Dynamical spin-1/2 model}
The 1D spin-1/2 model that we study here is dynamical, that is it has no random elements. The Hamiltonian is given by
\begin{eqnarray}
H &=&  \frac{J}{4}\sum_{s=1}^{L-1} \left( \sigma_s^x \sigma_{s+1}^x + \sigma_s^y \sigma_{s+1}^y +\Delta \sigma_s^z \sigma_{s+1}^z \right)  \\
 &+& \lambda \frac{J}{4} \sum_{s=1}^{L-2} \left( \sigma_s^x \sigma_{s+2}^x + \sigma_s^y \sigma_{s+2}^y +\Delta \sigma_s^z \sigma_{s+2}^z \right).
 \label{ham1} 
\end{eqnarray}
The first part of this Hamiltonian contains only nearest-neighbor couplings and it is associated with the mean field $H_0$. The second part describes next-nearest-neighbor couplings and represents the perturbation $V$. Differently from the previous model, $V$ is a local interaction in space. The Pauli matrices $\sigma^{x,y,z}_s$ act on site $s$; $L$ is the number of sites which is chosen even; the coupling constant $J=1$ sets the energy scale; $\Delta$ stands for the anisotropy of the interaction, and $\lambda$ is the ratio between  next-nearest-neighbor and nearest-neighbor couplings~\cite{Santos2009JMP,TorresKollmar2015}.  

The Hamiltonian conserves the total spin in the $z$-direction, ${\cal S}^z = \sum_{s=1}^L \sigma_s^z/2$. In what follows we consider the subspace ${\cal S}^z=-1$, which has $N=L/2-1$ excitations (up-spins) and dimension 
\[
{\cal D}=\dfrac{L!}{N!(L-N)!}.
\] 
The unperturbed Hamiltonian $H_0$ is integrable, but as $\lambda$ increases, $H$ crosses over to the chaotic regime~\cite{Santos2012PRL,Santos2012PRE}. For the parameters considered here, system size $L=16$, number of up-spins $N=7$, (so ${\cal D} = 11440$), anisotropy $\Delta=0.48$, and $\lambda=1$, the model is strongly chaotic in a large region of the spectrum. 

\subsection{Quench Dynamics}
To study the dynamics, we prepare the system in an unperturbed state  $\ket{k_0} $,
\begin{equation}
\ket{\psi(0)} =\ket{k_0}  = \sum_\alpha C_{k_0}^\alpha \ket{\alpha},
\end{equation}
where $C_{k_0}^\alpha =\langle \alpha|k_0 \rangle $ and $\ket{\alpha}$ are the exact energy eigenstates. The initial state $\ket{\psi(0)}$ evolves under the full Hamiltonian $H$ when the interaction $V$ is turned on. We consider initial states that have energy $E_{k_0}=\bra{k_0}H\ket{k_0}$  away from the edges of the spectrum of $H$. 

We notice that the initial state for the spin model is not a site-basis vector (computational basis vector) for which the spin on each site either points up or down in the $z$-direction, but it is instead an eigenstate of $H_0$. In analogy with the TBRE, we refer to these states as the unperturbed many-body basis.

The probability to find the evolved state in a basis state  $\ket{k}$ at the time $t$ is given by
\begin{eqnarray}
\label{Eq:Pk}
P_k(t) &=&  \left| \langle k |e^{-i Ht} |k_0 \rangle \right|^2 = |\langle k |\psi (t) \rangle |^2 \\
&& \nonumber \\
&=&\sum_{\alpha,\beta}  C_{k_0}^{\alpha *} C_{k}^{\alpha } C_{k_0}^{\beta } C_{k}^{\beta *} e^{-i(E^\beta-E^\alpha )t} .
\end{eqnarray}
The particular case where $k=k_0$ corresponds to the survival probability (also known as return probability), which can be written as
\begin{eqnarray}
P_{k_0}(t) &=& |\langle k_0 |\psi (t) \rangle |^2 = \left| \sum_{\alpha}  \left| C_{k_0}^{\alpha} \right|^2  e^{-i E^\alpha t} \right|^2 \nonumber \\
&=&  \left| \int dE\, e^{ - iEt} \rho_{k_0}(E) \right|^2,
\label{eq:fourier}
\end{eqnarray}
where 
\begin{equation}
\label{eq:ldos}
\rho_{k_0}(E)  \equiv \sum_{\alpha}  | C^{\alpha}_{k_0} |^2 \delta (E - E_\alpha )
\end{equation} 
is the LDOS, that is the energy distribution weighted by the components $| C^{\alpha}_{k_0} |^2$ of the initial state. The subscript $k_0$ in  Eq.~(\ref{eq:ldos}) stresses the important point that the LDOS depends  on the initial state $\ket{k_0}$. As evident from Eq.~(\ref{eq:fourier}), the survival probability is the Fourier transform of the LDOS. The inverse of the width $\Gamma$ of the LDOS gives the characteristic decay time of $P_{k_0}(t)$. 

The maximal size of the LDOS, obtained when $H_0$ is negligible and $H\sim V$, defines the energy shell, which is only a part of the total energy spectrum.
 The shape of the energy shell depends on the density of states, which in systems with few-body interactions typically has a Gaussian form~\cite{Brody1981}. The eigenstates of $H$ written in the unperturbed basis are chaotic when they fill the energy shell completely and the components $C_k^{\alpha}$ are random numbers following the Gaussian envelope of the energy shell~\cite{Santos2012PRL,Santos2012PRE}.

To quantify how the initial state spreads in time, in the many-body Hilbert space, we compute the number of principal components (also known as participation ratio),
\begin{equation}
N_{pc}  (t) = \frac{1}{\sum_k P_k(t)^2} = \frac{1}{\sum_k |\langle k |\psi (t) \rangle |^4}.
\label{Eq:PR}
\end{equation}
For the TBRE, we use the notation $\langle\langle N_{pc}  (t) \rangle\rangle$ to indicate average over the random configurations of the two-body interaction.

\section{OTOC for projection operators and number of principal components}
\label{Sec:PR-OTOC} 
The OTOC for two Hermitian operators ${\hat w}$ and ${\hat v}$ is defined as,
\begin{equation}
F_{v,w} (t) = \left<{\hat w}^{\dagger} (t){\hat v}(0)^{\dagger}{\hat w}(t){\hat v}(0)\right> 
\label{Eq:OTOC_f}
\end{equation}
where ${\hat w}(t)= e^{iHt} {\hat w}(0) e^{-iHt}$ is the operator in the Heisenberg representation. 
In the literature, $\langle.\rangle$ originally referred to the average over the canonical ensemble, but later, averages over all states of an unperturbed Hamiltonian or over one particular initial state $\ket{k_0}$, as we do here, have also been considered.

Written in terms of the initial state, the OTOC has a clear physical meaning, which can be explained as follows. Let us define the two states,
$$
\ket{ x(t)}  = {\hat w}(t) {\hat v}(0)\ket{k_0}
$$
and
$$
\ket{ y(t)}  = {\hat v}(0){\hat w}(t) \ket{k_0},
$$
which  represents the action of the two operators taken in the reversed order. The state $\ket{ x(t)}$ is obtained by first applying ${\hat v}$, then evolving forward with the full Hamiltonian for time $t$, applying  ${\hat w}$, and finally evolving backward for the same time $t$. For $\ket{y(t)}$, the order is exchanged: first the evolution is forward, then ${\hat w}$ is applied, followed by the backward evolution, and finally the application of ${\hat v}$. Thus, $F_{v,w}(t) $ quantifies the decay of the overlap between these two states, $\langle y(t) |x(t) \rangle$, caused by the exchanged action of the two operators  ${\hat v}(0)$ and ${\hat w}(t)$. It probes the way $ {\hat v }$ and  ${\hat w}$ inhibit the cancellation between forward  and backward  evolution. Equivalently, $F_{v,w}(t) $  measures the degree of non-commutativity between the two operators.
 
The OTOC is related to the $N_{pc}$ when in Eq.~(\ref{Eq:OTOC_f}) we use projection operators in the unperturbed many-body states, ${\hat w}(0) =\ket{k}\bra{k}$, ${\hat v}(0)=\ket{k'}\bra{k'}$, and compute the expectation value in the initial state $\ket{k_0}$. This gives,
\begin{equation}
\label{OTOC1}
\begin{array}{lll}
F_{k,k_0} (t)   &=   \bra{k_0} e^{iHt}  \ket{k}\bra{k}   e^{-iHt} \ket{k'} \times \\
&\\
& \bra{k'}  e^{iHt} \ket{k}\bra{k}   e^{-iHt}
  \ket{k'}\bra{k'}   k_0 \rangle  \\ &\\
& =   \bra{k_0} e^{iHt}\ket{k} \bra{k} e^{-iHt} \ket{k_0} \times \\ &\\
& \hspace{0.37 cm} \bra{k_0}e^{iHt} \ket{k} \bra{k} e^{-iHt}\ket{k_0} \\ &\\
&=  | \bra{k} e^{-iHt}\ket{k_0} |^4  &
\end{array}
\end{equation}
Since ${\hat v}(0)\ket{k_0} = \delta_{k',k_0} \ket{k'} $, it is clear that  to have a non-zero correlation function one needs to choose ${\hat v}(0)=\ket{k_0}\bra{k_0}$.
Comparing the equation above with Eq.~(\ref{Eq:PR}), one sees that
\begin{eqnarray}
\label{IPR-otoc1}
[N_{pc} (t) ]^{-1}    &=&      \sum_{k\ne k_0}  F_{k,k_0} (t) + F_{k_0,k_0} (t) \\
& = &O_{toc} (t) + P_{k_0}(t)^2  . \nonumber
\label{Eq:twoTerms}
\end{eqnarray}
In the above, we separate $k=k_0$ from $k\ne k_0$. We refer to $F_{k,k_0} (t) $ for $k \neq k_0$ as projection-OTOC's, while the autocorrelation function $F_{k_0,k_0} (t)= \bra{k_0}  e^{-iHt} \ket{k_0}|^4 = P_{k_0}(t)^2$ is simply the squared survival probability. We denote by $O_{toc} (t)$ the extensive sum over all projection-OTOC's,
\begin{equation}
O_{toc} (t) = \sum_{k\ne k_0} F_{k,{k_0}}(t).
\label{Eq:extensive}
\end{equation}
The inverse of the $N_{pc}$ is therefore $O_{toc} (t)$  plus the squared survival probability.

\section{Analytical estimates and numerical results}
\label{Sec:An-Num}

We now have the tools to compare the results for the $N_{pc}$ and the OTOC for the TBRE and the dynamical spin-1/2 model in the strongly chaotic regime. As mentioned above, the initial states have energy $E_{k_0}=\bra{k_0}H\ket{k_0}$ far from the edges of the spectrum. 

\subsection{TBRE: Number of principal components  and OTOC}

For the TBRE, we focus on initial states, where all particles are on a single level, which we choose to be the fifth level, such as in $\ket{0,0,0,0,N,0,0,0,0,0}$. States of this kind have $E_{k_0}$  close to the center of the band. This choice of initial state is made, because the number of directly coupled matrix elements is minimal and independent of $N$.  The number of states directly coupled with the initial state together with the strength of the perturbation determine the width of the LDOS and thus the decay rate of the survival probability.

\begin{figure}[t]
\vspace{0.cm}
\includegraphics[width=\columnwidth]{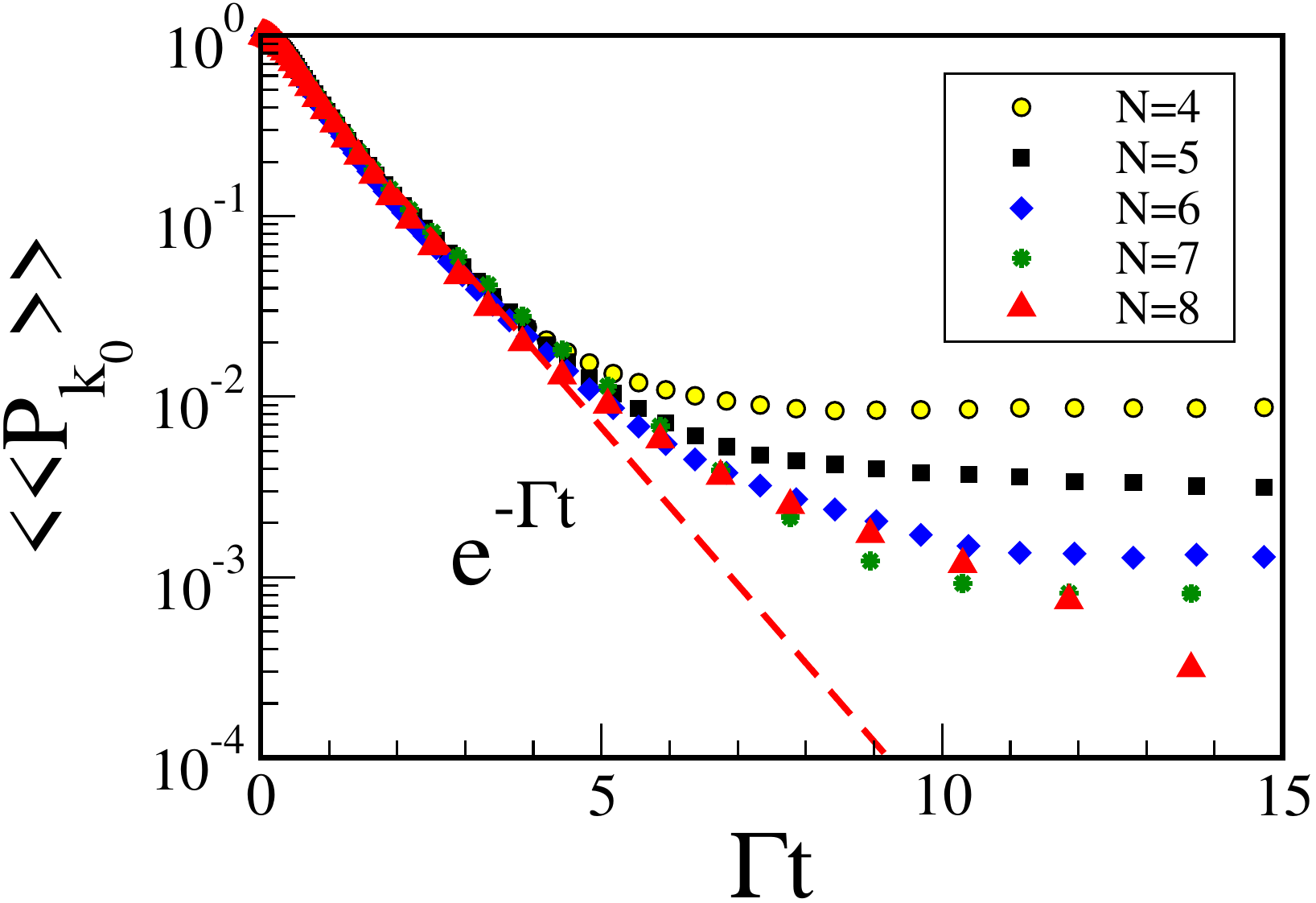}
\caption{Survival probability for the TBRE for initial states $\ket{k_0} = \ket{0,0,0,0,N,0,0,0,0,0,0} $ with different number of particles $N$, as indicated in the legend. The other parameters are $M=11$,  ${\cal V} = 0.4$. The dashed (red) line is the exponential fit for $N=8$ and $t<2$. The  exponential decay rate obtained from the fit is $\Gamma=2.4$. The numbers of random configurations chosen are $n_r=1000,500,100,50,5$ for $N=4,5,6,7,8$, respectively.
 }
\label{Fig:w0n}
\end{figure}

In Fig.~\ref{Fig:w0n}, we confirm that for the chosen perturbation and initial states, the survival probability decays exponentially and the decay rate is approximately independent of the number of particles. Needless to say, for very short time, $t\ll \Gamma^{-1}$,  the survival probability decays quadratically in time, as given by perturbation theory. This behavior is subsequently followed by a region of exponential decay with rate $\Gamma$, as seen in Fig.~\ref{Fig:w0n}.  This rate defines the timescale $t_\Gamma= 1/\Gamma $ for the depletion of the initial state~\cite{Borgonovi2019}. At this point, the probability to be in the initial state is reduced by a factor $1/e$. 

\subsubsection{Number of principal components}
\begin{figure}[t]
\vspace{0.cm}
\includegraphics[width=\columnwidth]{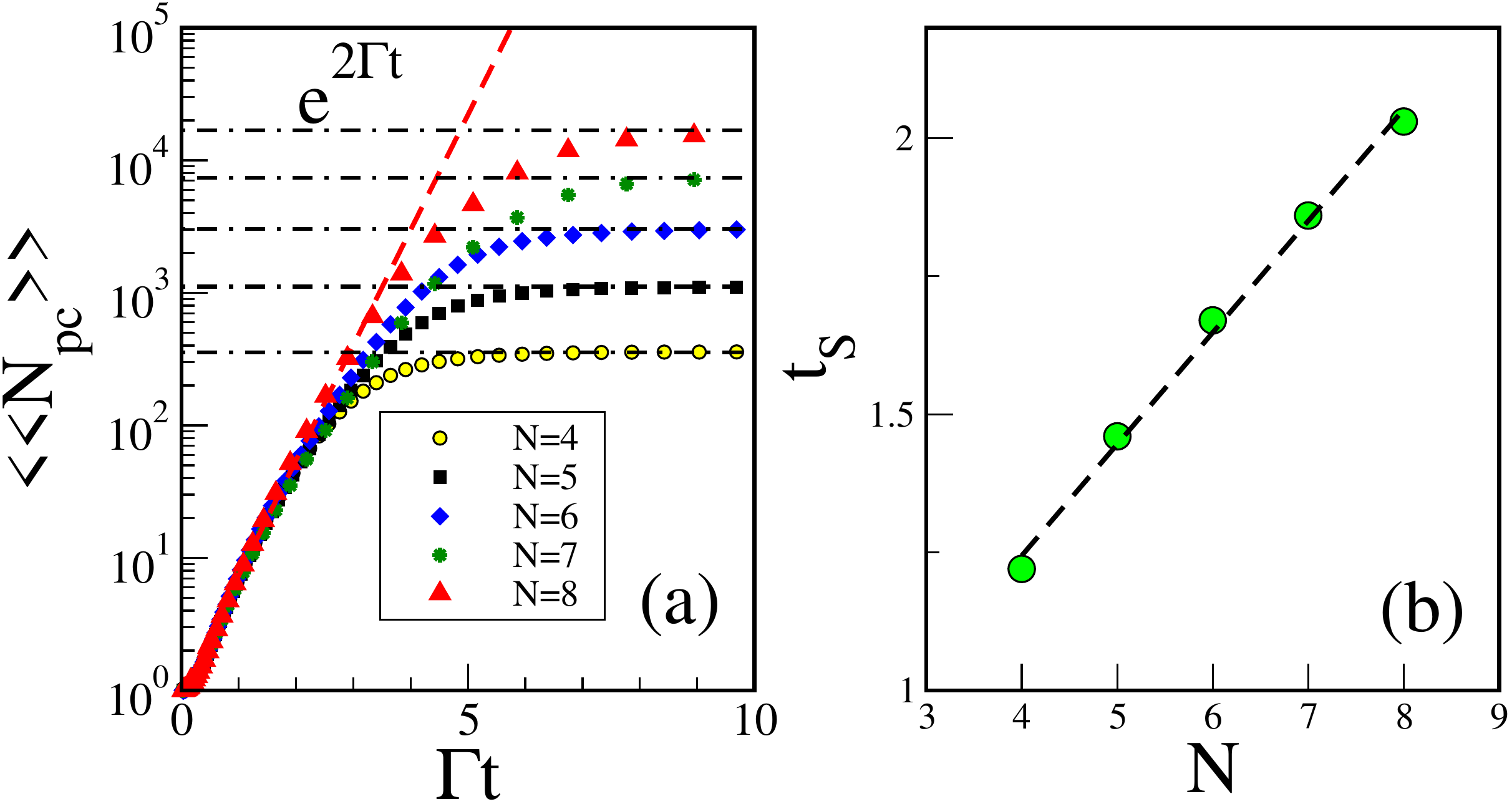}
\caption{(a) Growth in time of the number of principal components for the TBRE.  Different symbols stand for initial states with different numbers of particles $N$, as indicated in the legend. The horizontal lines represent  the saturation values $ \overline{N_{pc}^{\infty}}$. The dashed (red) line is the function $e^{2\Gamma t}$, where $\Gamma = 2.4$ was obtained in Fig.~\ref{Fig:w0n}. The horizontal dashed-dotted (black) lines indicate the asymptotic value $\overline{N_{pc}^{\infty}}$ given by Eq.~(\ref{t-ipr}).  
(b) Saturation times obtained by the intersection between the dashed (red) curve and the horizontal dashed-dotted (black) lines in panel (a),  as a function of the number of particles $N$. The dashed line is the best linear fit, $t_S \propto N$. The other parameters of the model are $M=11$,  ${\cal V} = 0.4$. The numbers of random configurations chosen are $n_r=1000,500,100,50,5$ for $N=4,5,6,7,8$, respectively. 
  }
\label{Fig:PR}
\end{figure}

The parameter $\Gamma$ is at the basis of a phenomenological cascade model~\cite{Borgonovi2019}, that describes in a coarse-grained way the spreading of the initial many-body state in the many-body Hilbert space.  The basic idea is to analyze the dynamics at different time steps, each being associated with the probability to find the  system in a specific subset of unperturbed many-body states, referred to as a ``class''. The class that contains only the initial state is the ${\cal M}_0 (k_0)$ class and the probability to be in this class is just the survival probability $P_{k_0} (t)$. ${\cal M}_1(k_0)$ is the set of all unperturbed states directly coupled to the initial state,
\[
{\cal M}_1(k_0) = \left\{ k\ne k_0, \ 1\leq k \leq {\cal D}, \ | \ \langle k | H | k_0 \rangle \ne 0 \right\}.
\]
The probability to be in this class is defined as  
\begin{equation}
W_1 (t) = \sum_{k\in {\cal M}_1(k_0) }|\langle k |\psi(t)\rangle |^2.
\end{equation}  
The subset with states coupled to $\ket{k_0}$ in second order of perturbation theory is ${\cal M}_2 (k_0)$, and so on. This description of the dynamics in terms of the spread of the wave packet in the many-body Hilbert space was also explored in~\cite{Altshuler1997,Flambaum2001b}. With this picture, we obtained in~\cite{Borgonovi2019}  approximate {\it rate equations} for the probability to find the system in each class. The sum of the square of these probabilities gives the
inverse of the number of principal components $N_{pc}$.  Our analysis predicted an exponential growth for $N_{pc}$ with exponent $2\Gamma$, which was verified numerically. This is shown in 
Fig.~\ref{Fig:PR}(a) for different initial states with increasing number of particles.

It is important to remark that the exponential increase of the number of principal components continues beyond $t_{\Gamma}$. At long times, since the many-body Hilbert space is finite, 
$N_{pc} (t)$ finally saturates to an equilibrium value, which is obtained by taking the infinite time average, 
\begin{eqnarray}
&&\left[\overline{N_{pc}^{\infty}}\right]^{-1} = \lim_{T\to \infty} \frac{1}{T} \int_0^T \ dt \sum_k |\langle k | e^{-iHt} |k_0 \rangle|^4 \nonumber \\
&& =  2 \sum_k \left(\sum_{\alpha}  |C_{k_0}^\alpha|^2 |C_{k}^\alpha|^2 \right)^2 - \sum_{\alpha}  | C_{k_0}^\alpha |^4 \sum_k |C_{k}^\alpha|^4 . \hspace{0.5 cm}
\label{t-ipr}
\end{eqnarray} 
An estimate of the saturation time $t_S$ can be obtained by equating $e^{2\Gamma t_S} \simeq \overline{N_{pc}^{\infty}}$. We showed in Ref.~\cite{Borgonovi2019} that for $M, N \gg 1$,  this estimate is given by $t_S \sim N t_\Gamma $. This result is seen clearly in Fig.~\ref{Fig:PR}~(b), together with a linear fit. The values for $t_S$ are obtained from the intersections in Fig.~\ref{Fig:PR}~(a) between the exponential curve and the horizontal lines, which indicate the saturation values from Eq.~(\ref{t-ipr}). We note that the saturation time $t_S$ was shown to coincide with the time necessary for the onset of the Bose-Einstein distribution for single-particle occupation numbers (for details see~\cite{Borgonovi2019b}). One can therefore identify $t_S$ with the thermalization time.

\subsubsection{Out-of-time ordered correlator}

\begin{figure}[t]
\vspace{0.cm}
\includegraphics[width=\columnwidth]{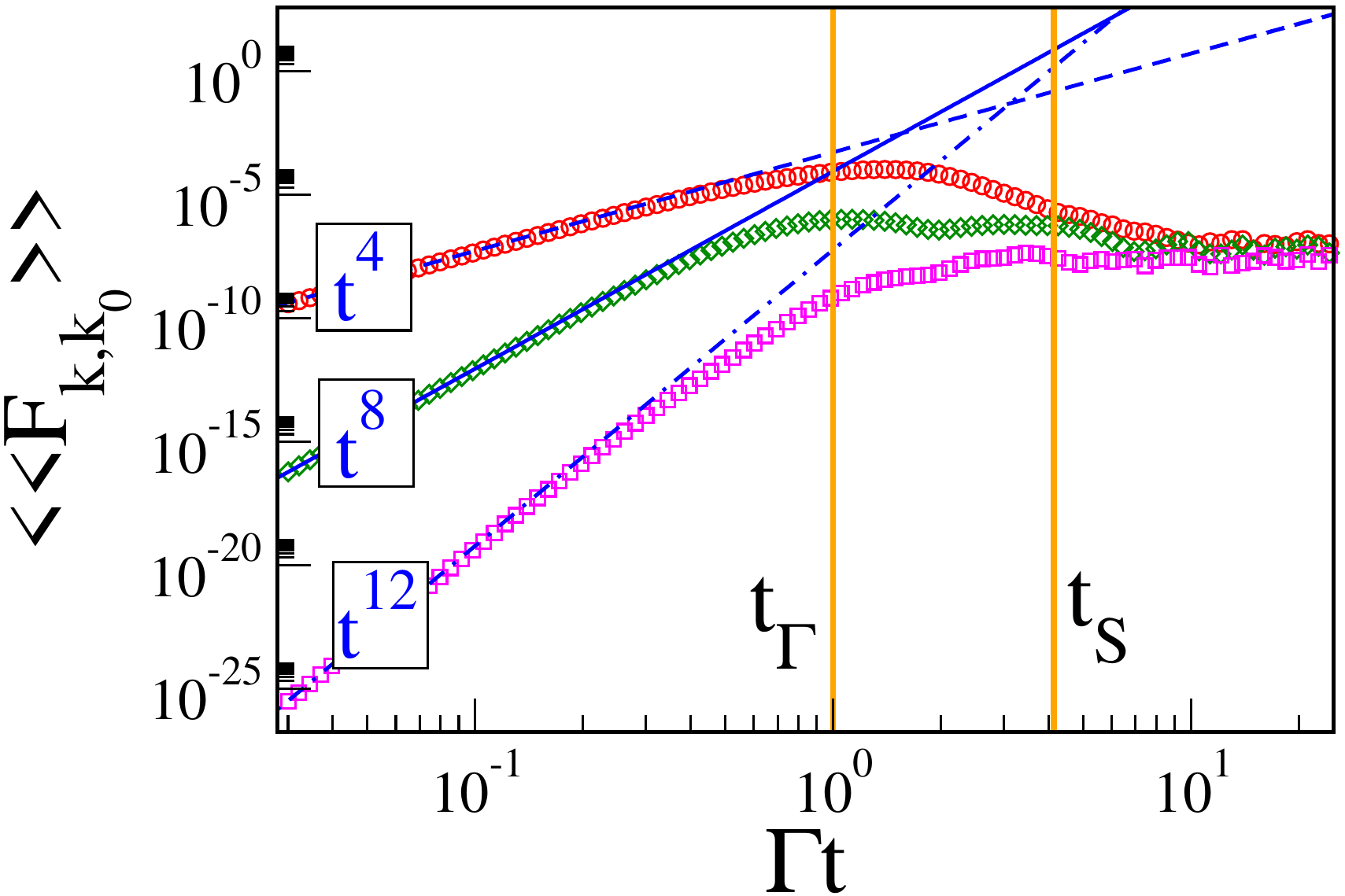}
\caption{OTOC's for projection operators with $k\neq k_0$ averaged over $100$ disorder realizations for the TBRE. From top to bottom,  $k$ in: $\langle k |H|k_0\rangle \ne 0$ (red); $\langle k |H|k_0\rangle = 0$ and  $\langle k |H^2|k_0\rangle \ne 0$ (green); and $\langle k |H|k_0\rangle = \langle k |H^2|k_0\rangle =  0$ (magenta). Dashed, solid, and dot-dashed lines  represent respectively the $t^4$, $t^8$ and $t^{12}$ behaviors.
 Vertical lines  indicate the depletion time $t_\Gamma$ and the thermalization time $t_S$.
 The initial state is chosen in the middle of the energy band and it has $6$ particles in the fifth single-particle energy level. The other parameters of the model are $M=11$ and $ {\cal V} = 0.4$.  }
\label{Fig:oto}
\end{figure}

We now proceed with the analysis of the OTOC and comparison with $N_{pc}$. The OTOC behavior  at short time can be obtained with the expansion,
\begin{eqnarray}
\hspace{-0.45 cm} F_{k,{k_0}}  (t) &=& | \langle k | e^{-iHt} | k_0\rangle |^4 \nonumber \\
&\simeq & | \delta_{k,k_0} -it H_{k,k_0} -\frac{1}{2} t^2 (H^2)_{k,k_0}+...|^4,
\label{o1}
\end{eqnarray}
where $H_{k,k_0} = \langle k | H | k_0 \rangle$.
For $k\ne k_0$, there are different behaviors, as listed below. 

(i) The first one corresponds to $k\in {\cal M}_1(k_0)$, for which one gets,
\begin{equation}
F_{k,{k_0}} (t) \simeq t^4 H_{k,k_0}^4 + o(t^6)
\quad {\rm for }  \quad k\in {\cal M}_1(k_0).
  \label{o1a}
\end{equation}
Taking the average over disorder realizations in the TBRE, we come to the following estimate,
\begin{eqnarray}
\langle \langle F_{k,{k_0}} (t)\rangle \rangle  &\simeq& t^4 \langle\langle H_{k,k_0}^4 \rangle\rangle \nonumber \\
& \simeq & 3 t^4 {\cal V}^4
\quad {\rm for }  \quad k\in {\cal M}_1(k_0).
  \label{o1ad}
\end{eqnarray}
To obtain the last line above, we took into account that $H_{k,k'}$ are Gaussian variables with  zero mean and variance ${\cal V}^2$.

(ii) For the case $k \in {\cal M}_2(k_0)$, one has a $t^8$ behavior,
\begin{equation}
F_{k,{k_0}} (t) \simeq \frac{1}{16} t^8 \left[ \sum_{k' \in {\cal M}_1} H_{k,k'} H_{k',k_0} \right]^4
\!\!\!\!\quad {\rm for }  \!\!\!\quad k\in {\cal M}_2(k_0).
  \label{o1b}
\end{equation}

(iii) For the projection-OTOC's of higher-order classes, where  $\langle k | H | k_0 \rangle = \langle k | H^2 | k_0 \rangle = 0 $, the initial numerical power-law growth  gives a  $t^{12}$ behavior. 

The behaviors $t^4$, $t^8$ and $t^{12}$ for the various projection-OTOC's are shown in Fig.~\ref{Fig:oto}, respectively as dashed, full and dot-dashed lines.  Perturbation theory is approximately valid for $t < t_\Gamma$. In the region marked by the exponential growth of the $N_{pc}$, that is $t_\Gamma < t < t_S$, the OTOC's have a non-generic and non-monotonous behavior. For $t > t_S$, the OTOC's  just show fluctuations around some equilibrium value.

\begin{figure}[t]
\vspace{0.cm}
\includegraphics[width=\columnwidth]{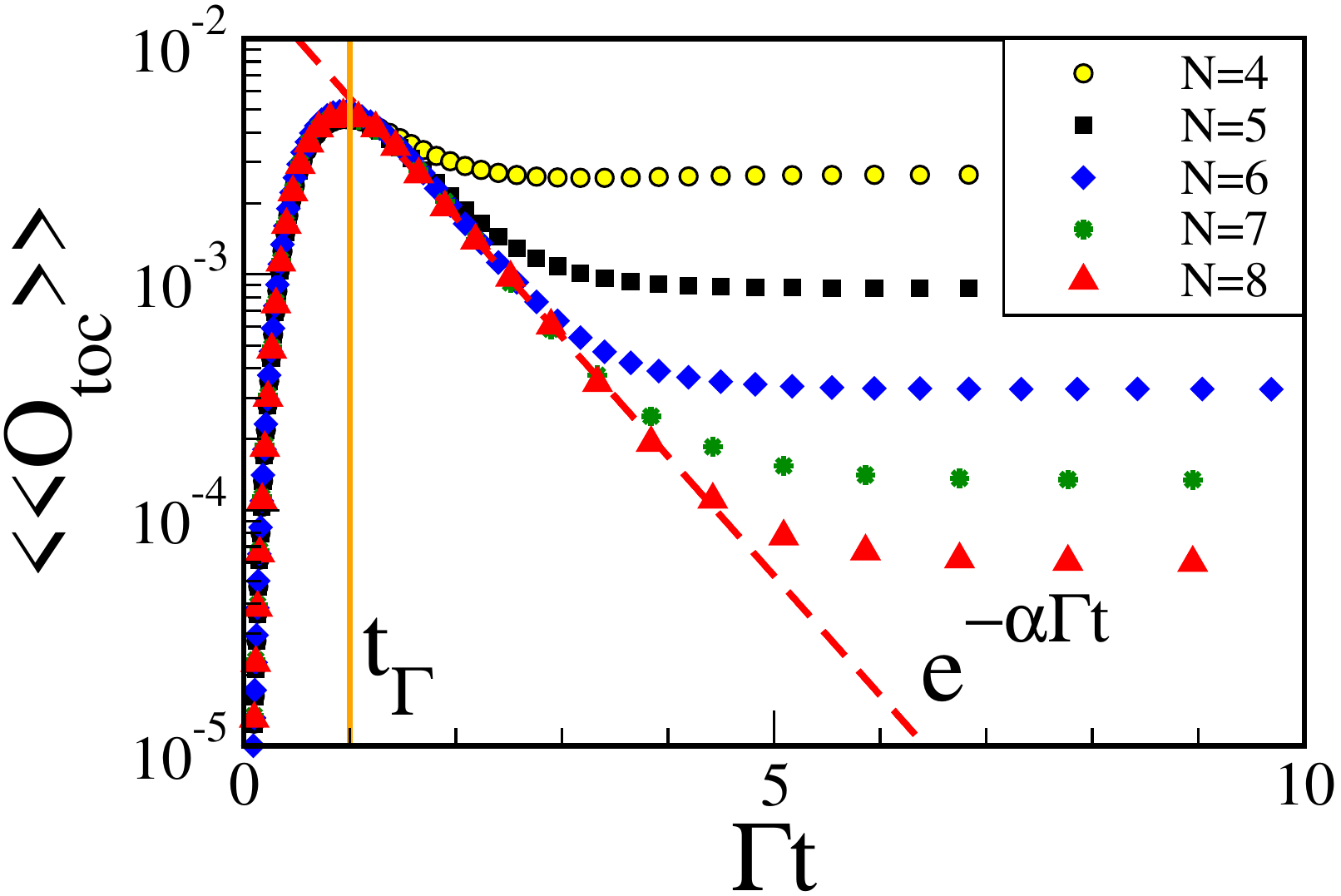}
\caption{Growth in time of the extensive sum of all projection-OTOC's. Different symbols stand  for initial states $\ket{k_0} = \ket{0,0,0,0,N,0,0,0,0,0,0} $ with different number of particles $N$, as indicated in the legend. The dashed (red) line is the fit with an exponential function $e^{\alpha\Gamma t}$ for the points with $N=8$ for $1.2 < \Gamma t < 4.5$. We fix  $\Gamma = 2.4$ (obtained from Fig.~\ref{Fig:w0n}) and get from the fitting $\alpha=1.2$. The other parameters of the model are $M=11$,  ${\cal V} = 0.4$. The number of random configurations chosen are $n_r=1000,500,100,50,5$ for $N=4,5,6,7,8$, respectively. 
}
\label{Fig:eo}
\end{figure}

In Fig.~\ref{Fig:eo}, we examine the behavior of the sum of all projection-OTOC's [Eq.~(\ref{Eq:extensive})]. Our figure shows the time dependence of the $\langle\langle O_{toc} (t)  \rangle\rangle$ for different numbers of particles. We can see that it reaches a maximum approximately at $t_\Gamma$ (vertical orange line),  when the probability to be in the initial state is reduced by a factor $1/e$. After this point, $\langle\langle O_{toc} (t)  \rangle\rangle$ decays exponentially, with an   exponent between $\Gamma$ and $2\Gamma$ (actually $1.2 \Gamma$ for this set of initial states). This exponent comes out from the sum of many different contributions from states belonging to different classes, and it cannot be obtained by taking into account the first-class states only. We note that extensive sums of local operators were also used in the analysis of the OTOC in Ref.~\cite{Kukuljan2017}, where it is argued that only the sum, and not a single local observable, can exhibit indefinite exponential growth in the thermodynamic limit.

We do not have yet a theory to extract the exponential decay rate for $ O_{toc} (t) $. It should be possible to associate the characteristic decay time for the sum $\sum_{k \in {\cal M}} F_{k,{k_0}} (t)$ of projection-OTOC's that belong to a specific class ${\cal M}$ to the scrambling time of the correlations during the flow from one class to the other. The timescale $t_S$ would emerge as a result of the summation of all different timescales associated to all classes. We leave this study to a future work. We note that the exponential decay of the out-of-time order correlators was recently obtained analytically for chaotic quantum maps~\cite{MataARXIV}. In that work, the  approach to the stationary value was found to occur with a rate determined by the Ruelle-Pollicot resonances.

\begin{figure}[t]
\vspace{0.cm}
\includegraphics[width=0.9\columnwidth]{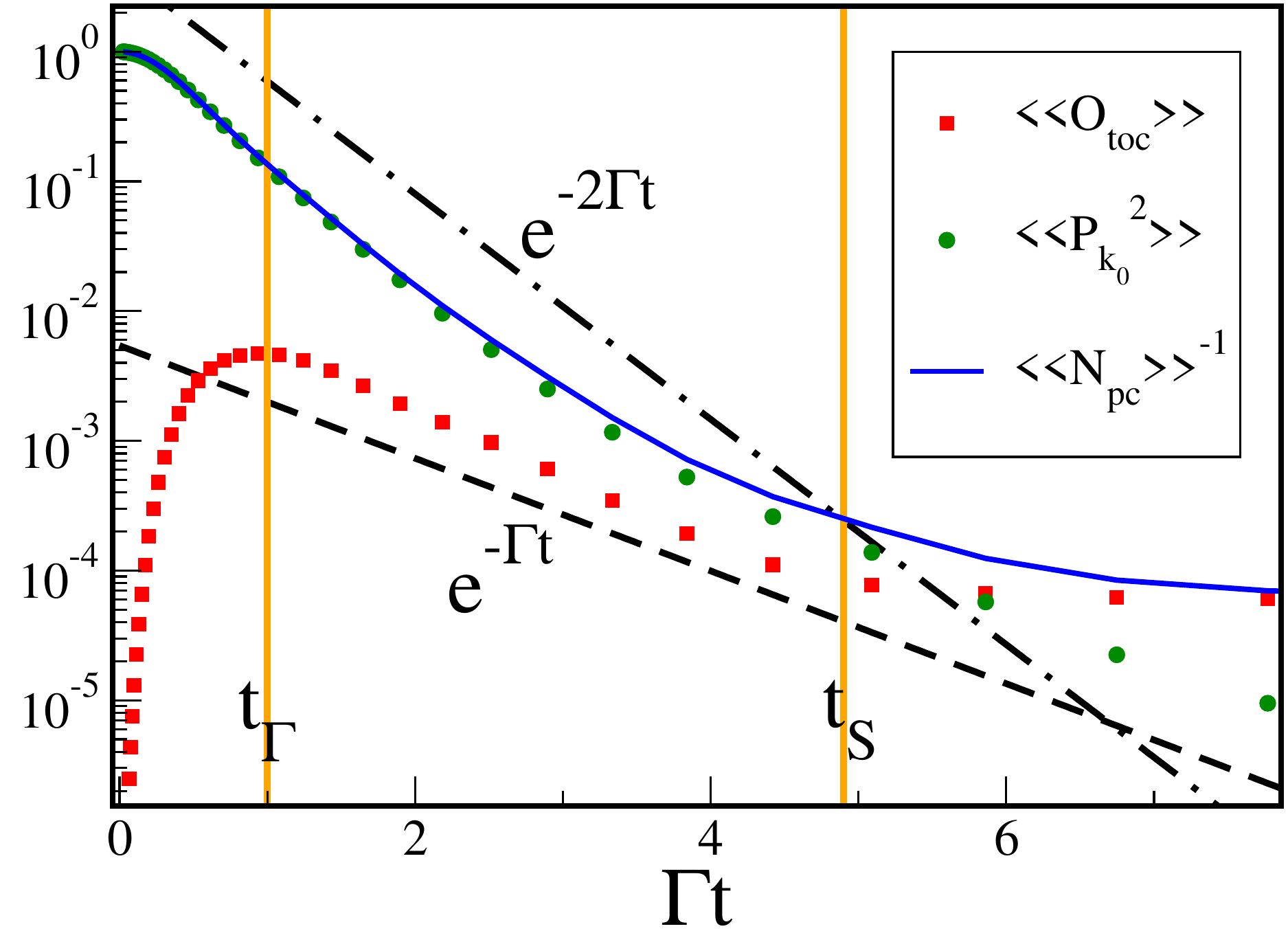}
\caption{Comparison between the sum of projection-OTOC's, the squared survival probability, and the inverse of the number of principal components, as indicated in the legend.
Vertical solid orange lines represent the depletion time $t_\Gamma$ and the saturation time $t_S$. The dashed and dashed-dotted lines stand for the $e^{-\Gamma t }$ and $e^{-2\Gamma t }$, respectively. The initial state, chosen in the middle of the energy band, has $8$ particles in the fifth single-particle energy level. The other parameters of the model are $M=11$, $ {\cal V} = 0.4$. The number of random configurations chosen is $n_r= 5 $.  
 }
\label{Fig:ow0}
\end{figure}

The exponential decay of $\langle\langle O_{toc} (t)  \rangle\rangle$ for $t_{\Gamma} < t < t_S$ indicates that the extensive sum of OTOC's plays an important role in the exponential growth of the number of principal components  beyond $t_{\Gamma}$.  In Fig.~\ref{Fig:ow0}, we compare the two terms appearing in the denominator of $N_{pc}$, that is $\langle\langle O_{toc} (t)  \rangle\rangle$ and $\langle \langle P_{k_0}(t)^2 \rangle\rangle$, for the case with $N=8$ particles.  Initially $\langle\langle N_{pc} (t)^{-1}  \rangle\rangle$ is entirely dominated by the squared survival probability. Later, due to the different decay rates for $\langle\langle O_{toc} (t)  \rangle\rangle$ and $\langle \langle P_{k_0}(t)^2 \rangle\rangle$, these two contributions become of the same order of magnitude and they eventually cross. 

As seen in Fig.~\ref{Fig:ow0}, for the system size and set of initial states considered, the crossing between $\langle\langle O_{toc} (t)  \rangle\rangle$ and $\langle \langle P_{k_0}(t)^2 \rangle\rangle$ occurs after the saturation time $t_S$. As a result,  the relaxation of $\langle\langle N_{pc} (t)  \rangle\rangle^{-1}$ to its infinite time-average value is entirely due to the saturation of $\langle\langle O_{toc} (t)  \rangle\rangle$. The two saturate roughly at the same time. In contrast, the squared survival probability reaches its stationary value at a timescale much larger than $t_S$. 

\begin{figure}[t]
\vspace{0.cm}
\includegraphics[width=\columnwidth]{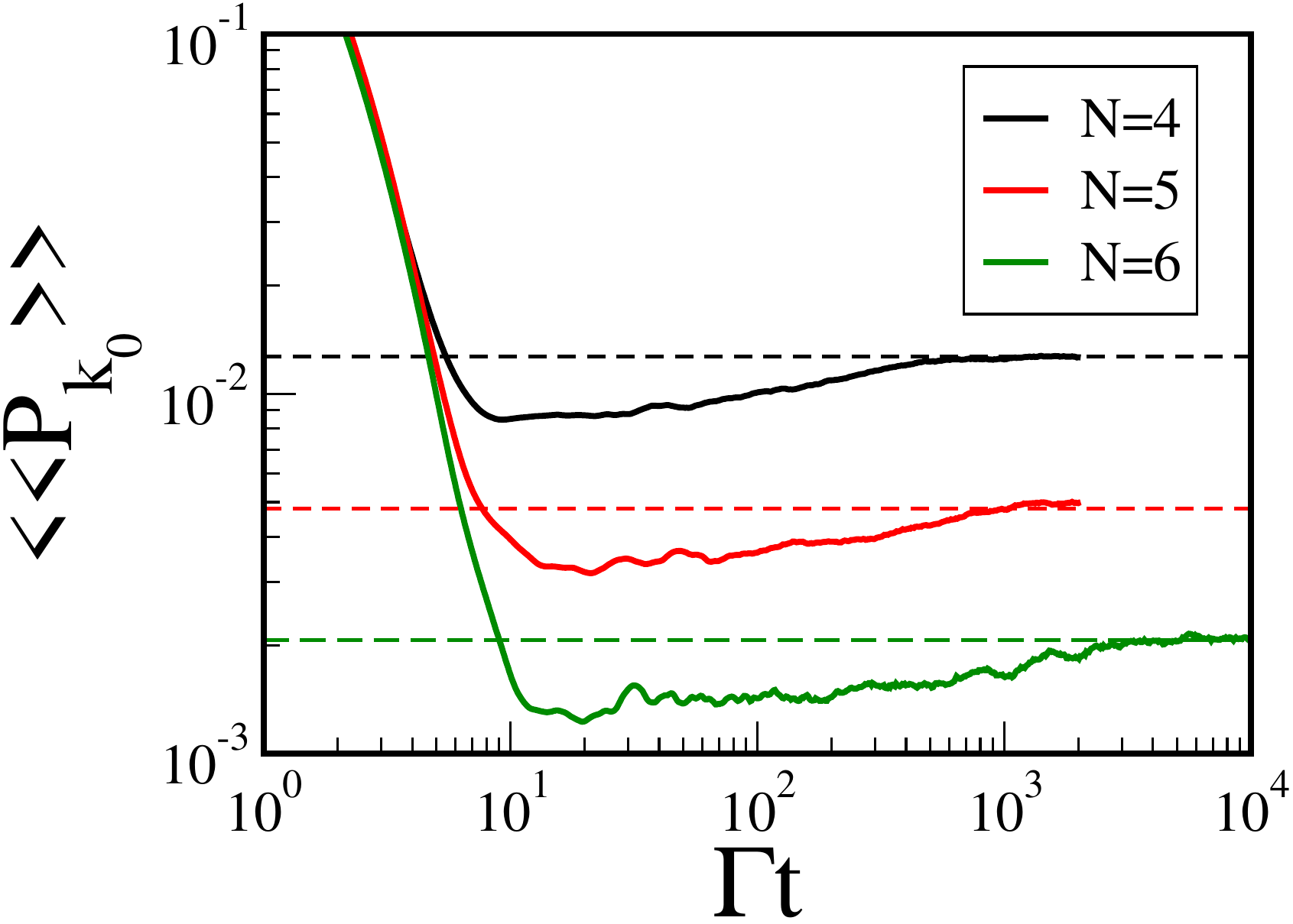}
\caption{Decay of the survival probability in time for TBRE.  The curves represent 3 initial states with $N=4,5,6$ particles in the fifth single-particle energy level.  The horizontal dashed lines represent the infinite time-average values, $\sum_{\alpha}|C_{k_0}^\alpha|^4$. The other parameters of the model are $M=11$ and ${\cal V} = 0.4$. The number of random configurations chosen is $n_r=1000, 500,100$ respectively.
}
\label{Fig:hole}
\end{figure}

Figure~\ref{Fig:hole}  illustrates the timescale for the relaxation of the survival probability. By comparing this time with the saturation time $t_S$ for $N_{pc}$ shown in  Fig.~\ref{Fig:PR}, we can see that the former is more that two orders of magnitude larger. This is due to the presence of the so-called correlation hole (see \cite{Torres2017,Torres2017PTR,Torres2018} and references therein), which is a dip below the saturation value. This hole is clearly visible for the survival probability, but it is not so evident for $\langle \langle N_{pc}(t) \rangle\rangle$ (for a comparison see Ref.~\cite{SchiulazARXIV}).  The correlation hole ends at the Heisenberg time, beyond which there are only fluctuations around the infinite-time average, given by $\sum_{\alpha}|C_{k_0}^\alpha|^4$.

\subsection{Spin-1/2 model: Number of principal components  and OTOC}
For the spin model, we do not perform any average, since the Hamiltonian has no random elements and a single initial state with energy $E_{k_0} \approx -0.5$ is considered. The results are very similar to those presented in Fig.~\ref{Fig:PR}, Fig.~\ref{Fig:oto}, and  Fig.~\ref{Fig:eo}.

\begin{figure}[t]
\hspace{0.38 cm}
\includegraphics[width=\columnwidth]{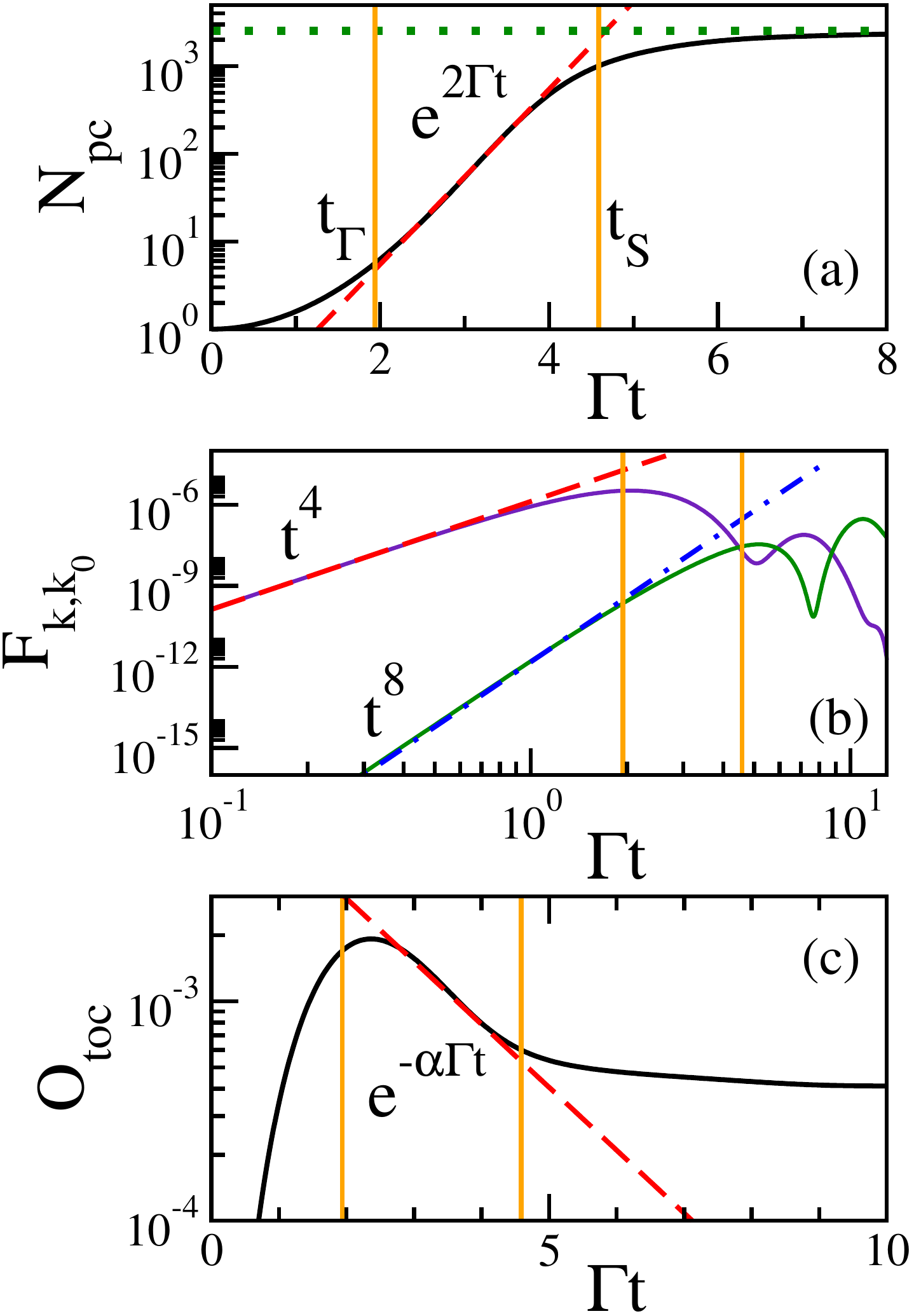}
\caption{Clean spin-1/2 model, $E_{k_0} \approx -0.5$. (a) Exponential growth in time of  the  number of principal components;  (b)  projection-OTOC's for some $k$'s; (c)  the extensive sum of projection-OTOC's. Vertical (orange) lines indicate $t_\Gamma$ and $t_S$.  In all panels, the numerical results are shown with solid curves. In (a), the dashed line indicates the exponential growth $e^{2\Gamma t}$ and the horizontal dotted line is for the infinite time average $\overline{N_{pc}^{\infty}}$. In (b) the dashed and dot-dashed curves represent the initial $t^4 $ and $t^8$ behavior for the probability to be in the first and second class, respectively. In (c) the dashed line is the exponential fitting $e^{-2\alpha\Gamma t}$ with  $\alpha=0.66$.
}
\label{Fig:PRspin}
\end{figure}

Figure~\ref{Fig:PRspin}~(a) shows the number of principal components, which grows as $e^{2\Gamma t}$ in the time interval $t_{\Gamma} < t < t_S$. In Fig.~\ref{Fig:PRspin}~(b), we depict the behavior of some projection-OTOC's. They show power-law growths proportional to $t^4$ and  $t^8$ for $t < t_\Gamma$, as seen also in Fig.~\ref{Fig:oto}. The behaviors become non-monotonic for $t_{\Gamma} < t < t_S$. From the figure, it is clear that states belonging to the first class (those having a $t^4$ initial growth) reach their maximal value before the states in the second class (those with a $t^8$ behavior). Since they reach the maximum at different times, they start to decay at different times, so we might expect a complicated behavior in the time region $t_\Gamma < t < t_S$. However, as clear from Fig.~\ref{Fig:PRspin}~(c), in this time interval, the extensive sum of all projection-OTOC's actually decays exponentially before saturation, with $\alpha=0.66$ in $e^{-\alpha\Gamma t}$. The result is similar to the one observed in Fig.~\ref{Fig:oto} for the TBRE.

The results for the spin model corroborate that for $t_{\Gamma} < t < t_S$, the sum given by $O_{toc} (t)$ contributes to the exponential behavior of $N_{pc}$, despite the fact that individually, the projection-OTOC's do not show any sign of exponential behavior in this time interval. We find a different decay exponent $\alpha$ from TBRE case. 
It is not clear at this point what this exponent might depend on, such as number of particles, energy of the initial state, and connectivity of the model. We leave this point for future investigations.

We notice that even though $H_0$ for the spin model can be solved with the Bethe ansatz, this is not at all trivial. Thus, we obtain numerically the eigenstates $\ket{k}$, used as the basis to write $H$. As a result, all matrix elements of $H$ become non-zero. To identify which elements correspond to effective couplings between the unperturbed states, we use a threshold $\xi=0.1$, that is, we assume that $\ket{k}$ is directly coupled with $\ket{k'}$ only if $H_{k,k'}> \xi |H_{k,k}-H_{k',k'}|$.

\section{Discussion}
\label{Sec:Conclusions}

We studied the relationship between the out-of-time ordered correlator (OTOC) and the number of principal components $N_{pc}$ (or participation ratio),  and their relevance to the relaxation process of many-body quantum systems. Two chaotic models were considered: One model belongs to the two-body random ensemble (TBRE), where randomness is introduced {\em ad hoc} as random couplings between many-body unperturbed states, and the other is a clean system of spin-$1/2$ particles on a linear chain with non-random two-body interactions. 

In a recent work~\cite{Borgonovi2019}, we had shown that, starting with a single many-body state of the unperturbed Hamiltonian $H_0$, the effective number of unperturbed many-body states participating in the dynamics, dictated by the perturbed Hamiltonian $H=H_0+V$, increases exponentially in time. This happens when the inter-particle interactions are sufficiently strong and the many-body eigenstates are superpositions of many effectively pseudo-random components, which is a main feature of strong quantum chaos. The quantity employed to characterize the spread of the initial wave packet in the Hilbert space was the number of principal components $N_{pc}$.

For strong perturbation, namely $H_0\sim V$, we found that $N_{pc}(t)$ increases as $e^{2 \Gamma t}$, where $\Gamma$ is the width of the LDOS. Our numerical data, as well as the analytical estimates, showed that this exponential behavior holds up to the saturation time $t_S \sim N t_{\Gamma}$, where $N$ is the number of particles for the TBRE and number of excitations for the spin model. This timescale is larger than the time $t_{\Gamma} \sim 1/{\Gamma}$ for the effective decrease of the survival probability.

In the present paper, we showed that $N_{pc}$ is the square of the survival probability plus the sum of all projection-OTOC's. For the latter, the operators ${\hat w}$ and ${\hat v}$ in Eq.~(\ref{Eq:OTOC_f}) are projection operators in the many-body Hilbert space, ${\hat w}$ being the projection on a state other than the initial state.

Our semi-analytical description of $N_{pc}(t)$  was based on the spread of the initial wave packet into different classes of unperturbed many-body states. At the shortest timescale, only the many-body states of $H_0$ directly coupled to the initial state by the two-body interactions get excited. Later in time, the wave packet propagates to those states which are coupled to the initial state in the second order of perturbation theory, and even later, higher orders are reached successively. This dynamics may be compared with the spread (mixing) of packets of classical trajectories in phase space: initially the whole phase space is scarcely occupied, but as time grows it gets more densely occupied. Within this picture, the projection-OTOC's describe the flow of the wave packet probability between specific classes. At short time, each one increases as $t^4, t^8, t^{12}$, depending on the class the OTOC is associated with, and in accordance with our analytical estimates. After reaching a maximal value, the projection-OTOC's  decay to a stationary value given by the infinite-time average value. In the course of this process, none of the individual projection-OTOC's shows an exponential behavior. It is only the  sum of the projection-OTOC's over all classes that decays exponentially for $t> t_{\Gamma}$.  This non-monotonic  behavior contrasts with that for the autocorrelation function (squared survival probability), which decays as $e^{-2 \Gamma t}$ already at short times.

It should be possible to associate to the sum of the projection-OTOC's belonging to a specific class ${\cal M}$, a characteristic decay time that represents the scrambling time for that class. The saturation of the entire dynamics at $t_S$ happens after the saturation of the projection-OTOC's for all classes.  After the time $t_S \sim N t_\Gamma $, the system is fully equilibrated (thermalized) in a finite but very large domain of the unperturbed basis. 

We finish this conclusion with a discussion about the quantum-classical correspondence for chaotic many-body systems. For this, we recall that the LDOS, which has width $\Gamma$,  has a well defined classical limit  with width $\Gamma_{cl}$~\cite{Borgonovi1998b,Luna2000,Izrailev2001,Luna2001,Luna2002}. Our results show that for  $N_{pc} (t)$, which is a global observable,  the timescale $t_S$ over which one can speak of exponential instability diverges in the thermodynamic limit, provided the semiclassical limit $\Gamma \to \Gamma_{cl}$ is done before  $N \rightarrow \infty$. This suggests that {\em there may be global observables for which the quantum-classical correspondence remains indefinitely in the thermodynamic limit.} 
 
The divergence of $t_S$ does not contradict the conventional picture of the Ehrenfest theorem, according to which the timescale of the quantum-classical correspondence for one-body chaotic systems is very small, $t_E \sim \ln(1/\hbar)$. As shown for the KR model, this is the timescale for a {\it local observable}, but there is another timescale, $t_D \sim 1/\hbar$, corresponding to the dynamical localization in the momentum space, which is related to a {\it global observable}. Therefore, the timescales for the quantum-classical correspondence depend on the choice of the observable and can vary significantly from one observable to another. Our study for many-body models focused on the global observable $N_{pc}$, rather than on local observables. There is not yet any direct comparison between $N_{pc}$ and a classical analog. We suggested in Ref.~\cite{Borgonovi2019} that such comparison will have to be done with the use of the Kolmogorov-Sinai entropy, which is the main characteristic of the dynamics for classical many-body systems, whose dynamics occurs in a $2 N$ dimensional phase space. 

One should mention that the {\it quantum diffusion} in the KR is not a ``true'' diffusion as that occurring in classical systems. As shown in~\cite{Shepelyansky1983}, the quantum diffusion is completely reversible, despite the presence of small, but finite errors associated with any numerical calculation. This is at variance  with classical diffusion, which is non-reversible due to the exponential sensitivity with respect to unavoidable computation errors. This is a distinctive property of the observed quantum-classical correspondence for the wave packet width in the momentum space. One can conjecture that a similar picture should arise for many-body chaos. Even though the quantum-classical correspondence may look very good for global observables (for the number of principal components in our case), quantum properties such as local quantum correlations and entanglement may still be present during the relaxation process and even at thermalization. In fact, it was recently shown numerically and semi-analytically in Ref.~\cite{Borgonovi2019b} that the Bose-Einstein distribution for occupation numbers emerges on the same timescale as the thermalization time $t_S$. This implies the {\it coexistence of classical and quantum features} in the dynamics  on a very large timescale 
 $t >  t_S \sim N/\Gamma $. The quantum-classical correspondence for many-body systems is a challenging problem that requires further studies.

\newpage
\begin{acknowledgments}
F.B. acknowledges support by the I.S. INFNDynSysMath. F.M.I. acknowledges financial support from CONACyT (Grant No. 286633). L.F.S. is supported by the U.S. National Science Foundation (NSF) Grant No.~DMR-1603418. 
\end{acknowledgments}

\bibliography{biblioOTOC_PR}

\begin{thebibliography}{72}%
\makeatletter
\providecommand \@ifxundefined [1]{%
 \@ifx{#1\undefined}
}%
\providecommand \@ifnum [1]{%
 \ifnum #1\expandafter \@firstoftwo
 \else \expandafter \@secondoftwo
 \fi
}%
\providecommand \@ifx [1]{%
 \ifx #1\expandafter \@firstoftwo
 \else \expandafter \@secondoftwo
 \fi
}%
\providecommand \natexlab [1]{#1}%
\providecommand \enquote  [1]{``#1''}%
\providecommand \bibnamefont  [1]{#1}%
\providecommand \bibfnamefont [1]{#1}%
\providecommand \citenamefont [1]{#1}%
\providecommand \href@noop [0]{\@secondoftwo}%
\providecommand \href [0]{\begingroup \@sanitize@url \@href}%
\providecommand \@href[1]{\@@startlink{#1}\@@href}%
\providecommand \@@href[1]{\endgroup#1\@@endlink}%
\providecommand \@sanitize@url [0]{\catcode `\\12\catcode `\$12\catcode
  `\&12\catcode `\#12\catcode `\^12\catcode `\_12\catcode `\%12\relax}%
\providecommand \@@startlink[1]{}%
\providecommand \@@endlink[0]{}%
\providecommand \url  [0]{\begingroup\@sanitize@url \@url }%
\providecommand \@url [1]{\endgroup\@href {#1}{\urlprefix }}%
\providecommand \urlprefix  [0]{URL }%
\providecommand \Eprint [0]{\href }%
\providecommand \doibase [0]{http://dx.doi.org/}%
\providecommand \selectlanguage [0]{\@gobble}%
\providecommand \bibinfo  [0]{\@secondoftwo}%
\providecommand \bibfield  [0]{\@secondoftwo}%
\providecommand \translation [1]{[#1]}%
\providecommand \BibitemOpen [0]{}%
\providecommand \bibitemStop [0]{}%
\providecommand \bibitemNoStop [0]{.\EOS\space}%
\providecommand \EOS [0]{\spacefactor3000\relax}%
\providecommand \BibitemShut  [1]{\csname bibitem#1\endcsname}%
\let\auto@bib@innerbib\@empty
\bibitem [{\citenamefont {Kaufman}\ \emph {et~al.}(2016)\citenamefont
  {Kaufman}, \citenamefont {M.~Eric~Tai}, \citenamefont {Rispoli},
  \citenamefont {Schittko}, \citenamefont {Preiss},\ and\ \citenamefont
  {Greiner}}]{Kaufman2016}%
  \BibitemOpen
  \bibfield  {author} {\bibinfo {author} {\bibfnamefont {Adam~M.}\ \bibnamefont
  {Kaufman}}, \bibinfo {author} {\bibfnamefont {Alexander~Lukin}\ \bibnamefont
  {M.~Eric~Tai}}, \bibinfo {author} {\bibfnamefont {Matthew}\ \bibnamefont
  {Rispoli}}, \bibinfo {author} {\bibfnamefont {Robert}\ \bibnamefont
  {Schittko}}, \bibinfo {author} {\bibfnamefont {Philipp~M.}\ \bibnamefont
  {Preiss}}, \ and\ \bibinfo {author} {\bibfnamefont {Markus}\ \bibnamefont
  {Greiner}},\ }\bibfield  {title} {\enquote {\bibinfo {title} {Quantum
  thermalization through entanglement in an isolated many-body system},}\
  }\href {\doibase 10.1126/science.aaf6725} {\bibfield  {journal} {\bibinfo
  {journal} {Science}\ }\textbf {\bibinfo {volume} {353}},\ \bibinfo {pages}
  {794} (\bibinfo {year} {2016})}\BibitemShut {NoStop}%
\bibitem [{\citenamefont {Lewis-Swan}\ \emph {et~al.}()\citenamefont
  {Lewis-Swan}, \citenamefont {Safavi-Naini}, \citenamefont {Bollinger},\ and\
  \citenamefont {Rey}}]{LewisARXIV}%
  \BibitemOpen
  \bibfield  {author} {\bibinfo {author} {\bibfnamefont {R.~J.}\ \bibnamefont
  {Lewis-Swan}}, \bibinfo {author} {\bibfnamefont {A.}~\bibnamefont
  {Safavi-Naini}}, \bibinfo {author} {\bibfnamefont {J.~J.}\ \bibnamefont
  {Bollinger}}, \ and\ \bibinfo {author} {\bibfnamefont {A.~M.}\ \bibnamefont
  {Rey}},\ }\href@noop {} {\enquote {\bibinfo {title} {Unifying fast
  scrambling, thermalization and entanglement through the measurement of
  {FOTOC}s in the {D}icke model},}\ }\bibinfo {note}
  {ArXiv:1808.07134}\BibitemShut {NoStop}%
\bibitem [{\citenamefont {Wei}\ \emph {et~al.}()\citenamefont {Wei},
  \citenamefont {Peng}, \citenamefont {Shtanko}, \citenamefont {Marvian},
  \citenamefont {Lloyd}, \citenamefont {Ramanathan},\ and\ \citenamefont
  {Cappellaro}}]{WeiARXIV}%
  \BibitemOpen
  \bibfield  {author} {\bibinfo {author} {\bibfnamefont {Ken~Xuan}\
  \bibnamefont {Wei}}, \bibinfo {author} {\bibfnamefont {Pai}\ \bibnamefont
  {Peng}}, \bibinfo {author} {\bibfnamefont {Oles}\ \bibnamefont {Shtanko}},
  \bibinfo {author} {\bibfnamefont {Iman}\ \bibnamefont {Marvian}}, \bibinfo
  {author} {\bibfnamefont {Seth}\ \bibnamefont {Lloyd}}, \bibinfo {author}
  {\bibfnamefont {Chandrasekhar}\ \bibnamefont {Ramanathan}}, \ and\ \bibinfo
  {author} {\bibfnamefont {Paola}\ \bibnamefont {Cappellaro}},\ }\href@noop {}
  {\enquote {\bibinfo {title} {Emergent prethermalization signatures in
  out-of-time ordered correlations},}\ }\bibinfo {note}
  {ArXiv:1812.04776}\BibitemShut {NoStop}%
\bibitem [{\citenamefont {Borgonovi}\ \emph {et~al.}(2016)\citenamefont
  {Borgonovi}, \citenamefont {Izrailev}, \citenamefont {Santos},\ and\
  \citenamefont {Zelevinsky}}]{Borgonovi2016}%
  \BibitemOpen
  \bibfield  {author} {\bibinfo {author} {\bibfnamefont {F.}~\bibnamefont
  {Borgonovi}}, \bibinfo {author} {\bibfnamefont {F.~M.}\ \bibnamefont
  {Izrailev}}, \bibinfo {author} {\bibfnamefont {L.~F.}\ \bibnamefont
  {Santos}}, \ and\ \bibinfo {author} {\bibfnamefont {V.~G.}\ \bibnamefont
  {Zelevinsky}},\ }\bibfield  {title} {\enquote {\bibinfo {title} {Quantum
  chaos and thermalization in isolated systems of interacting particles},}\
  }\href {\doibase 10.1016/j.physrep.2016.02.005} {\bibfield  {journal}
  {\bibinfo  {journal} {Phys. Rep.}\ }\textbf {\bibinfo {volume} {626}},\
  \bibinfo {pages} {1} (\bibinfo {year} {2016})}\BibitemShut {NoStop}%
\bibitem [{\citenamefont {D'Alessio}\ \emph {et~al.}(2016)\citenamefont
  {D'Alessio}, \citenamefont {Kafri}, \citenamefont {Polkovnikov},\ and\
  \citenamefont {Rigol}}]{Alessio2016}%
  \BibitemOpen
  \bibfield  {author} {\bibinfo {author} {\bibfnamefont {Luca}\ \bibnamefont
  {D'Alessio}}, \bibinfo {author} {\bibfnamefont {Yariv}\ \bibnamefont
  {Kafri}}, \bibinfo {author} {\bibfnamefont {Anatoli}\ \bibnamefont
  {Polkovnikov}}, \ and\ \bibinfo {author} {\bibfnamefont {Marcos}\
  \bibnamefont {Rigol}},\ }\bibfield  {title} {\enquote {\bibinfo {title} {From
  quantum chaos and eigenstate thermalization to statistical mechanics and
  thermodynamics},}\ }\href {\doibase 10.1080/00018732.2016.1198134} {\bibfield
   {journal} {\bibinfo  {journal} {Adv. in Phys.}\ }\textbf {\bibinfo {volume}
  {65}},\ \bibinfo {pages} {239--362} (\bibinfo {year} {2016})}\BibitemShut
  {NoStop}%
\bibitem [{\citenamefont {Borgonovi}\ \emph {et~al.}(2019)\citenamefont
  {Borgonovi}, \citenamefont {Izrailev},\ and\ \citenamefont
  {Santos}}]{Borgonovi2019}%
  \BibitemOpen
  \bibfield  {author} {\bibinfo {author} {\bibfnamefont {Fausto}\ \bibnamefont
  {Borgonovi}}, \bibinfo {author} {\bibfnamefont {Felix~M.}\ \bibnamefont
  {Izrailev}}, \ and\ \bibinfo {author} {\bibfnamefont {Lea~F.}\ \bibnamefont
  {Santos}},\ }\bibfield  {title} {\enquote {\bibinfo {title} {Exponentially
  fast dynamics of chaotic many-body systems},}\ }\href {\doibase
  10.1103/PhysRevE.99.010101} {\bibfield  {journal} {\bibinfo  {journal} {Phys.
  Rev. E}\ }\textbf {\bibinfo {volume} {99}},\ \bibinfo {pages} {010101}
  (\bibinfo {year} {2019})}\BibitemShut {NoStop}%
\bibitem [{\citenamefont {Schiulaz}\ \emph {et~al.}()\citenamefont {Schiulaz},
  \citenamefont {Torres-Herrera},\ and\ \citenamefont
  {Santos}}]{SchiulazARXIV}%
  \BibitemOpen
  \bibfield  {author} {\bibinfo {author} {\bibfnamefont {M.}~\bibnamefont
  {Schiulaz}}, \bibinfo {author} {\bibfnamefont {E.~J.}\ \bibnamefont
  {Torres-Herrera}}, \ and\ \bibinfo {author} {\bibfnamefont {Lea~F.}\
  \bibnamefont {Santos}},\ }\href@noop {} {\enquote {\bibinfo {title} {Thouless
  and relaxation time scales in many-body quantum systems},}\ }\bibinfo {note}
  {ArXiv:1807.07577}\BibitemShut {NoStop}%
\bibitem [{\citenamefont {.}()}]{DymarskyARXIVThouless}%
  \BibitemOpen
  \bibfield  {author} {\bibinfo {author} {\bibfnamefont {A.~Dymarsky}\
  \bibnamefont {.}},\ }\href@noop {} {\enquote {\bibinfo {title} {Mechanism of
  slow equilibration of isolated quantum systems},}\ }\bibinfo {note}
  {ArXiv:1806.04187}\BibitemShut {NoStop}%
\bibitem [{\citenamefont {Eriksson}\ \emph {et~al.}(2018)\citenamefont
  {Eriksson}, \citenamefont {Bengtsson}, \citenamefont {Karabulut},
  \citenamefont {Kavoulakis},\ and\ \citenamefont {Reimann}}]{Eriksson2018}%
  \BibitemOpen
  \bibfield  {author} {\bibinfo {author} {\bibfnamefont {G}~\bibnamefont
  {Eriksson}}, \bibinfo {author} {\bibfnamefont {J}~\bibnamefont {Bengtsson}},
  \bibinfo {author} {\bibfnamefont {E~\~A}\ \bibnamefont {Karabulut}}, \bibinfo
  {author} {\bibfnamefont {G~M}\ \bibnamefont {Kavoulakis}}, \ and\ \bibinfo
  {author} {\bibfnamefont {S~M}\ \bibnamefont {Reimann}},\ }\bibfield  {title}
  {\enquote {\bibinfo {title} {Finite-size effects in the dynamics of few
  bosons in a ring potential},}\ }\href
  {http://stacks.iop.org/0953-4075/51/i=3/a=035504} {\bibfield  {journal}
  {\bibinfo  {journal} {J. Phys. B}\ }\textbf {\bibinfo {volume} {51}},\
  \bibinfo {pages} {035504} (\bibinfo {year} {2018})}\BibitemShut {NoStop}%
\bibitem [{\citenamefont {Maldacena}\ \emph {et~al.}(2017)\citenamefont
  {Maldacena}, \citenamefont {Shenker},\ and\ \citenamefont
  {Z.}}]{Maldacena2017FP}%
  \BibitemOpen
  \bibfield  {author} {\bibinfo {author} {\bibfnamefont {J}~\bibnamefont
  {Maldacena}}, \bibinfo {author} {\bibfnamefont {S.~H.}\ \bibnamefont
  {Shenker}}, \ and\ \bibinfo {author} {\bibfnamefont {Yang}\ \bibnamefont
  {Z.}},\ }\bibfield  {title} {\enquote {\bibinfo {title} {Diving into
  traversable wormholes},}\ }\href@noop {} {\bibfield  {journal} {\bibinfo
  {journal} {Fortschr. Phys.}\ }\textbf {\bibinfo {volume} {65}},\ \bibinfo
  {pages} {1700034} (\bibinfo {year} {2017})}\BibitemShut {NoStop}%
\bibitem [{Lar()}]{Larkin1969}%
  \BibitemOpen
  \href@noop {} {}\bibinfo {note} {A. Larkin and Yu. N. Ovchinnikov, Zh. Eksp.
  Teor. Fiz. 55, 2262 (1969) [``Quasiclassical Method in the Theory of
  Superconductivity'', Sov. Phys. JETP 28, 1200 (1969)].}\BibitemShut {Stop}%
\bibitem [{\citenamefont {Sekino}\ and\ \citenamefont
  {Susskind}(2008)}]{Sekino2008}%
  \BibitemOpen
  \bibfield  {author} {\bibinfo {author} {\bibfnamefont {Yasuhiro}\
  \bibnamefont {Sekino}}\ and\ \bibinfo {author} {\bibfnamefont
  {L.}~\bibnamefont {Susskind}},\ }\bibfield  {title} {\enquote {\bibinfo
  {title} {Fast scramblers},}\ }\href
  {http://stacks.iop.org/1126-6708/2008/i=10/a=065} {\bibfield  {journal}
  {\bibinfo  {journal} {J. High Energy Physics}\ }\textbf {\bibinfo {volume}
  {2008}},\ \bibinfo {pages} {065} (\bibinfo {year} {2008})}\BibitemShut
  {NoStop}%
\bibitem [{\citenamefont {Kitaev}({\natexlab{a}})}]{KitaevTALK}%
  \BibitemOpen
  \bibfield  {author} {\bibinfo {author} {\bibfnamefont {A.}~\bibnamefont
  {Kitaev}},\ }\href@noop {} {\enquote {\bibinfo {title} {Hidden correlations
  in the {H}awking radiation and thermal noise, talk at breakthrough physics
  prize symposium, nov. 10, 2014},}\ } ({\natexlab{a}}),\ \bibinfo {note}
  {https://www.youtube.com/watch?v=OQ9qN8j7EZI}\BibitemShut {NoStop}%
\bibitem [{\citenamefont {Maldacena}\ and\ \citenamefont
  {Stanford}(2016)}]{Maldacena2016PRD}%
  \BibitemOpen
  \bibfield  {author} {\bibinfo {author} {\bibfnamefont {Juan}\ \bibnamefont
  {Maldacena}}\ and\ \bibinfo {author} {\bibfnamefont {Douglas}\ \bibnamefont
  {Stanford}},\ }\bibfield  {title} {\enquote {\bibinfo {title} {Remarks on the
  {S}achdev-{Y}e-{K}itaev model},}\ }\href {\doibase
  10.1103/PhysRevD.94.106002} {\bibfield  {journal} {\bibinfo  {journal} {Phys.
  Rev. D}\ }\textbf {\bibinfo {volume} {94}},\ \bibinfo {pages} {106002}
  (\bibinfo {year} {2016})}\BibitemShut {NoStop}%
\bibitem [{\citenamefont {Maldacena}\ \emph {et~al.}(2016)\citenamefont
  {Maldacena}, \citenamefont {Shenker},\ and\ \citenamefont
  {Stanford}}]{Maldacena2016JHEP}%
  \BibitemOpen
  \bibfield  {author} {\bibinfo {author} {\bibfnamefont {Juan}\ \bibnamefont
  {Maldacena}}, \bibinfo {author} {\bibfnamefont {Stephen~H.}\ \bibnamefont
  {Shenker}}, \ and\ \bibinfo {author} {\bibfnamefont {Douglas}\ \bibnamefont
  {Stanford}},\ }\bibfield  {title} {\enquote {\bibinfo {title} {A bound on
  chaos},}\ }\href {\doibase 10.1007/JHEP08(2016)106} {\bibfield  {journal}
  {\bibinfo  {journal} {J. High Energy Phys.}\ }\textbf {\bibinfo {volume}
  {2016}},\ \bibinfo {pages} {106} (\bibinfo {year} {2016})}\BibitemShut
  {NoStop}%
\bibitem [{\citenamefont {G\"arttner}\ \emph {et~al.}(2017)\citenamefont
  {G\"arttner}, \citenamefont {Bohnet}, \citenamefont {Safavi-Naini},
  \citenamefont {Wall}, \citenamefont {Bollinger},\ and\ \citenamefont
  {Rey}}]{Garttner2017}%
  \BibitemOpen
  \bibfield  {author} {\bibinfo {author} {\bibfnamefont {Martin}\ \bibnamefont
  {G\"arttner}}, \bibinfo {author} {\bibfnamefont {Justin~G.}\ \bibnamefont
  {Bohnet}}, \bibinfo {author} {\bibfnamefont {Arghavan}\ \bibnamefont
  {Safavi-Naini}}, \bibinfo {author} {\bibfnamefont {Michael~L.}\ \bibnamefont
  {Wall}}, \bibinfo {author} {\bibfnamefont {John~J.}\ \bibnamefont
  {Bollinger}}, \ and\ \bibinfo {author} {\bibfnamefont {Ana~Maria}\
  \bibnamefont {Rey}},\ }\bibfield  {title} {\enquote {\bibinfo {title}
  {Measuring out-of-time-order correlations and multiple quantum spectra in a
  trapped-ion quantum magnet},}\ }\href {http://dx.doi.org/10.1038/nphys4119}
  {\bibfield  {journal} {\bibinfo  {journal} {Nat. Phys.}\ }\textbf {\bibinfo
  {volume} {13}},\ \bibinfo {pages} {781 -- 786} (\bibinfo {year}
  {2017})}\BibitemShut {NoStop}%
\bibitem [{\citenamefont {Li}\ \emph {et~al.}(2017)\citenamefont {Li},
  \citenamefont {Fan}, \citenamefont {Wang}, \citenamefont {Ye}, \citenamefont
  {Zeng}, \citenamefont {Zhai}, \citenamefont {Peng},\ and\ \citenamefont
  {Du}}]{Li2017}%
  \BibitemOpen
  \bibfield  {author} {\bibinfo {author} {\bibfnamefont {Jun}\ \bibnamefont
  {Li}}, \bibinfo {author} {\bibfnamefont {Ruihua}\ \bibnamefont {Fan}},
  \bibinfo {author} {\bibfnamefont {Hengyan}\ \bibnamefont {Wang}}, \bibinfo
  {author} {\bibfnamefont {Bingtian}\ \bibnamefont {Ye}}, \bibinfo {author}
  {\bibfnamefont {Bei}\ \bibnamefont {Zeng}}, \bibinfo {author} {\bibfnamefont
  {Hui}\ \bibnamefont {Zhai}}, \bibinfo {author} {\bibfnamefont {Xinhua}\
  \bibnamefont {Peng}}, \ and\ \bibinfo {author} {\bibfnamefont {Jiangfeng}\
  \bibnamefont {Du}},\ }\bibfield  {title} {\enquote {\bibinfo {title}
  {Measuring out-of-time-order correlators on a nuclear magnetic resonance
  quantum simulator},}\ }\href {\doibase 10.1103/PhysRevX.7.031011} {\bibfield
  {journal} {\bibinfo  {journal} {Phys. Rev. X}\ }\textbf {\bibinfo {volume}
  {7}},\ \bibinfo {pages} {031011} (\bibinfo {year} {2017})}\BibitemShut
  {NoStop}%
\bibitem [{\citenamefont {Niknam}\ \emph {et~al.}()\citenamefont {Niknam},
  \citenamefont {Santos},\ and\ \citenamefont {Cory}}]{NiknamARXIV}%
  \BibitemOpen
  \bibfield  {author} {\bibinfo {author} {\bibfnamefont {Mohamad}\ \bibnamefont
  {Niknam}}, \bibinfo {author} {\bibfnamefont {Lea~F.}\ \bibnamefont {Santos}},
  \ and\ \bibinfo {author} {\bibfnamefont {David~G.}\ \bibnamefont {Cory}},\
  }\href@noop {} {\enquote {\bibinfo {title} {Sensitivity of quantum
  information to environment perturbations \\measured with the
  out-of-time-order correlation function},}\ }\bibinfo {note}
  {ArXiv:1808.04375}\BibitemShut {NoStop}%
\bibitem [{\citenamefont {Swingle}(2018)}]{Swingle2018}%
  \BibitemOpen
  \bibfield  {author} {\bibinfo {author} {\bibfnamefont {Brian}\ \bibnamefont
  {Swingle}},\ }\bibfield  {title} {\enquote {\bibinfo {title} {Unscrambling
  the physics of out-of-time-order correlators},}\ }\href {\doibase
  10.1038/s41567-018-0295-5} {\bibfield  {journal} {\bibinfo  {journal} {Nature
  Physics}\ }\textbf {\bibinfo {volume} {14}},\ \bibinfo {pages} {988}
  (\bibinfo {year} {2018})}\BibitemShut {NoStop}%
\bibitem [{\citenamefont {Rammensee}\ \emph {et~al.}(2018)\citenamefont
  {Rammensee}, \citenamefont {Urbina},\ and\ \citenamefont
  {Richter}}]{Rammensee2018}%
  \BibitemOpen
  \bibfield  {author} {\bibinfo {author} {\bibfnamefont {Josef}\ \bibnamefont
  {Rammensee}}, \bibinfo {author} {\bibfnamefont {Juan~Diego}\ \bibnamefont
  {Urbina}}, \ and\ \bibinfo {author} {\bibfnamefont {Klaus}\ \bibnamefont
  {Richter}},\ }\bibfield  {title} {\enquote {\bibinfo {title} {Many-body
  quantum interference and the saturation of out-of-time-order correlators},}\
  }\href {\doibase 10.1103/PhysRevLett.121.124101} {\bibfield  {journal}
  {\bibinfo  {journal} {Phys. Rev. Lett.}\ }\textbf {\bibinfo {volume} {121}},\
  \bibinfo {pages} {124101} (\bibinfo {year} {2018})}\BibitemShut {NoStop}%
\bibitem [{\citenamefont {Rodolfo A.~Jalabert}(2018)}]{Jalabert2018}%
  \BibitemOpen
  \bibfield  {author} {\bibinfo {author} {\bibfnamefont {Diego A.~Wisniacki}\
  \bibnamefont {Rodolfo A.~Jalabert}, \bibfnamefont {Ignacio Garcia-Mata}},\
  }\bibfield  {title} {\enquote {\bibinfo {title} {Semiclassical theory of
  out-of-time-order correlators for low-dimensional classically chaotic
  systems},}\ }\href@noop {} {\bibfield  {journal} {\bibinfo  {journal} {Phys.
  Rev. E}\ }\textbf {\bibinfo {volume} {98}},\ \bibinfo {pages} {062218}
  (\bibinfo {year} {2018})}\BibitemShut {NoStop}%
\bibitem [{\citenamefont {Peres}(1996)}]{Peres1996}%
  \BibitemOpen
  \bibfield  {author} {\bibinfo {author} {\bibfnamefont {Asher}\ \bibnamefont
  {Peres}},\ }\bibfield  {title} {\enquote {\bibinfo {title} {Chaotic evolution
  in quantum mechanics},}\ }\href {\doibase 10.1103/PhysRevE.53.4524}
  {\bibfield  {journal} {\bibinfo  {journal} {Phys. Rev. E}\ }\textbf {\bibinfo
  {volume} {53}},\ \bibinfo {pages} {4524--4527} (\bibinfo {year}
  {1996})}\BibitemShut {NoStop}%
\bibitem [{\citenamefont {Levstein}\ \emph {et~al.}(1998)\citenamefont
  {Levstein}, \citenamefont {Usaj},\ and\ \citenamefont
  {Pastawski}}]{Levstein1998}%
  \BibitemOpen
  \bibfield  {author} {\bibinfo {author} {\bibfnamefont {Patricia~R.}\
  \bibnamefont {Levstein}}, \bibinfo {author} {\bibfnamefont {Gonzalo}\
  \bibnamefont {Usaj}}, \ and\ \bibinfo {author} {\bibfnamefont {Horacio~M.}\
  \bibnamefont {Pastawski}},\ }\bibfield  {title} {\enquote {\bibinfo {title}
  {Attenuation of polarization echoes in nuclear magnetic resonance: A study of
  the emergence of dynamical irreversibility in many-body quantum systems},}\
  }\href {\doibase 10.1063/1.475664} {\bibfield  {journal} {\bibinfo  {journal}
  {J. Chem. Phys.}\ }\textbf {\bibinfo {volume} {108}},\ \bibinfo {pages}
  {2718--2724} (\bibinfo {year} {1998})}\BibitemShut {NoStop}%
\bibitem [{\citenamefont {Cucchietti}\ \emph {et~al.}(2002)\citenamefont
  {Cucchietti}, \citenamefont {Lewenkopf}, \citenamefont {Mucciolo},
  \citenamefont {Pastawski},\ and\ \citenamefont {Vallejos}}]{Cucchietti2002}%
  \BibitemOpen
  \bibfield  {author} {\bibinfo {author} {\bibfnamefont {F.~M.}\ \bibnamefont
  {Cucchietti}}, \bibinfo {author} {\bibfnamefont {C.~H.}\ \bibnamefont
  {Lewenkopf}}, \bibinfo {author} {\bibfnamefont {E.~R.}\ \bibnamefont
  {Mucciolo}}, \bibinfo {author} {\bibfnamefont {H.~M.}\ \bibnamefont
  {Pastawski}}, \ and\ \bibinfo {author} {\bibfnamefont {R.~O.}\ \bibnamefont
  {Vallejos}},\ }\bibfield  {title} {\enquote {\bibinfo {title} {Measuring the
  {L}yapunov exponent using quantum mechanics},}\ }\href@noop {} {\bibfield
  {journal} {\bibinfo  {journal} {Phys. Rev. E}\ }\textbf {\bibinfo {volume}
  {65}},\ \bibinfo {pages} {046209} (\bibinfo {year} {2002})}\BibitemShut
  {NoStop}%
\bibitem [{\citenamefont {Gorin}\ \emph {et~al.}(2006)\citenamefont {Gorin},
  \citenamefont {Prosen}, \citenamefont {Seligman},\ and\ \citenamefont
  {\ifmmode \check{Z}\else \v{Z}\fi{}nidari\ifmmode~\check{c}\else
  \v{c}\fi{}}}]{Gorin2006}%
  \BibitemOpen
  \bibfield  {author} {\bibinfo {author} {\bibfnamefont {T.}~\bibnamefont
  {Gorin}}, \bibinfo {author} {\bibfnamefont {Tomaz}\ \bibnamefont {Prosen}},
  \bibinfo {author} {\bibfnamefont {Thomas~H.}\ \bibnamefont {Seligman}}, \
  and\ \bibinfo {author} {\bibfnamefont {Marko}\ \bibnamefont {\ifmmode
  \check{Z}\else \v{Z}\fi{}nidari\ifmmode~\check{c}\else \v{c}\fi{}}},\
  }\bibfield  {title} {\enquote {\bibinfo {title} {Dynamics of loschmidt echoes
  and fidelity decay},}\ }\href@noop {} {\bibfield  {journal} {\bibinfo
  {journal} {Phys. Rep.}\ }\textbf {\bibinfo {volume} {435}},\ \bibinfo {pages}
  {33 -- 156} (\bibinfo {year} {2006})}\BibitemShut {NoStop}%
\bibitem [{\citenamefont {Elsayed}\ and\ \citenamefont
  {Fine}(2015)}]{Elsayed2015}%
  \BibitemOpen
  \bibfield  {author} {\bibinfo {author} {\bibfnamefont {Tarek~A}\ \bibnamefont
  {Elsayed}}\ and\ \bibinfo {author} {\bibfnamefont {Boris~V}\ \bibnamefont
  {Fine}},\ }\bibfield  {title} {\enquote {\bibinfo {title} {Sensitivity to
  small perturbations in systems of large quantum spins},}\ }\href
  {http://stacks.iop.org/1402-4896/2015/i=T165/a=014011} {\bibfield  {journal}
  {\bibinfo  {journal} {Phys. Scr.}\ }\textbf {\bibinfo {volume} {2015}},\
  \bibinfo {pages} {014011} (\bibinfo {year} {2015})}\BibitemShut {NoStop}%
\bibitem [{\citenamefont {Rozenbaum}\ \emph {et~al.}(2017)\citenamefont
  {Rozenbaum}, \citenamefont {Ganeshan},\ and\ \citenamefont
  {Galitski}}]{Rozenbaum2017}%
  \BibitemOpen
  \bibfield  {author} {\bibinfo {author} {\bibfnamefont {Efim~B.}\ \bibnamefont
  {Rozenbaum}}, \bibinfo {author} {\bibfnamefont {Sriram}\ \bibnamefont
  {Ganeshan}}, \ and\ \bibinfo {author} {\bibfnamefont {Victor}\ \bibnamefont
  {Galitski}},\ }\bibfield  {title} {\enquote {\bibinfo {title} {{L}yapunov
  exponent and out-of-time-ordered correlator's growth rate in a chaotic
  system},}\ }\href {\doibase 10.1103/PhysRevLett.118.086801} {\bibfield
  {journal} {\bibinfo  {journal} {Phys. Rev. Lett.}\ }\textbf {\bibinfo
  {volume} {118}},\ \bibinfo {pages} {086801} (\bibinfo {year}
  {2017})}\BibitemShut {NoStop}%
\bibitem [{\citenamefont {Rozenbaum}\ \emph {et~al.}()\citenamefont
  {Rozenbaum}, \citenamefont {Ganeshan},\ and\ \citenamefont
  {Galitski}}]{RozenbaumARXIV}%
  \BibitemOpen
  \bibfield  {author} {\bibinfo {author} {\bibfnamefont {E.~B.}\ \bibnamefont
  {Rozenbaum}}, \bibinfo {author} {\bibfnamefont {S.}~\bibnamefont {Ganeshan}},
  \ and\ \bibinfo {author} {\bibfnamefont {V.}~\bibnamefont {Galitski}},\
  }\href@noop {} {\enquote {\bibinfo {title} {Universal level statistics of the
  out-of-time-ordered operator},}\ }\bibinfo {note}
  {ArXiv:1801.10591}\BibitemShut {NoStop}%
\bibitem [{\citenamefont {Ch\'avez-Carlos}\ \emph {et~al.}(2019)\citenamefont
  {Ch\'avez-Carlos}, \citenamefont {L\'opez-del Carpio}, \citenamefont
  {Bastarrachea-Magnani}, \citenamefont {Str\'ansk\'y}, \citenamefont
  {Lerma-Hern\'andez}, \citenamefont {Santos},\ and\ \citenamefont
  {Hirsch}}]{Chavez2019}%
  \BibitemOpen
  \bibfield  {author} {\bibinfo {author} {\bibfnamefont {Jorge}\ \bibnamefont
  {Ch\'avez-Carlos}}, \bibinfo {author} {\bibfnamefont {B.}~\bibnamefont
  {L\'opez-del Carpio}}, \bibinfo {author} {\bibfnamefont {Miguel~A.}\
  \bibnamefont {Bastarrachea-Magnani}}, \bibinfo {author} {\bibfnamefont
  {Pavel}\ \bibnamefont {Str\'ansk\'y}}, \bibinfo {author} {\bibfnamefont
  {Sergio}\ \bibnamefont {Lerma-Hern\'andez}}, \bibinfo {author} {\bibfnamefont
  {Lea~F.}\ \bibnamefont {Santos}}, \ and\ \bibinfo {author} {\bibfnamefont
  {Jorge~G.}\ \bibnamefont {Hirsch}},\ }\bibfield  {title} {\enquote {\bibinfo
  {title} {Quantum and classical lyapunov exponents in atom-field interaction
  systems},}\ }\href {\doibase 10.1103/PhysRevLett.122.024101} {\bibfield
  {journal} {\bibinfo  {journal} {Phys. Rev. Lett.}\ }\textbf {\bibinfo
  {volume} {122}},\ \bibinfo {pages} {024101} (\bibinfo {year}
  {2019})}\BibitemShut {NoStop}%
\bibitem [{\citenamefont {Casati}\ \emph {et~al.}(1979)\citenamefont {Casati},
  \citenamefont {Chirikov}, \citenamefont {Izrailev},\ and\ \citenamefont
  {Ford}}]{Casati1979}%
  \BibitemOpen
  \bibfield  {author} {\bibinfo {author} {\bibfnamefont {G.}~\bibnamefont
  {Casati}}, \bibinfo {author} {\bibfnamefont {B.~V.}\ \bibnamefont
  {Chirikov}}, \bibinfo {author} {\bibfnamefont {F.~M.}\ \bibnamefont
  {Izrailev}}, \ and\ \bibinfo {author} {\bibfnamefont {J.}~\bibnamefont
  {Ford}},\ }\bibfield  {title} {\enquote {\bibinfo {title} {Stochastic
  behavior of a quantum pendulum under a periodic perturbation},}\ }\href@noop
  {} {\bibfield  {journal} {\bibinfo  {journal} {Lect. Notes in Phys.}\
  }\textbf {\bibinfo {volume} {9399}},\ \bibinfo {pages} {334--352} (\bibinfo
  {year} {1979})}\BibitemShut {NoStop}%
\bibitem [{\citenamefont {Chirikov}\ \emph {et~al.}(1981)\citenamefont
  {Chirikov}, \citenamefont {Izrailev},\ and\ \citenamefont
  {Shepelyansky}}]{Chirikov1981}%
  \BibitemOpen
  \bibfield  {author} {\bibinfo {author} {\bibfnamefont {Boris~V.}\
  \bibnamefont {Chirikov}}, \bibinfo {author} {\bibfnamefont {F.~M.}\
  \bibnamefont {Izrailev}}, \ and\ \bibinfo {author} {\bibfnamefont {D.~L.}\
  \bibnamefont {Shepelyansky}},\ }\href@noop {} {\bibfield  {journal} {\bibinfo
   {journal} {Sov. Sci. Rev. C}\ }\textbf {\bibinfo {volume} {2}},\ \bibinfo
  {pages} {209} (\bibinfo {year} {1981})}\BibitemShut {NoStop}%
\bibitem [{\citenamefont {Shepelyansky}(1983)}]{Shepelyansky1983}%
  \BibitemOpen
  \bibfield  {author} {\bibinfo {author} {\bibfnamefont {D.~L.}\ \bibnamefont
  {Shepelyansky}},\ }\bibfield  {title} {\enquote {\bibinfo {title} {Some
  statistical properties of simple classically stochastic quantum systems},}\
  }\href@noop {} {\bibfield  {journal} {\bibinfo  {journal} {Physica D}\
  }\textbf {\bibinfo {volume} {8}},\ \bibinfo {pages} {208} (\bibinfo {year}
  {1983})}\BibitemShut {NoStop}%
\bibitem [{\citenamefont {Berman}\ and\ \citenamefont
  {Zaslavsky}(1978)}]{Berman1978}%
  \BibitemOpen
  \bibfield  {author} {\bibinfo {author} {\bibfnamefont {G.~P.}\ \bibnamefont
  {Berman}}\ and\ \bibinfo {author} {\bibfnamefont {G.~M.}\ \bibnamefont
  {Zaslavsky}},\ }\bibfield  {title} {\enquote {\bibinfo {title} {Condition of
  stochasticity in quantum nonlinear systems},}\ }\href@noop {} {\bibfield
  {journal} {\bibinfo  {journal} {Physica A}\ }\textbf {\bibinfo {volume}
  {91}},\ \bibinfo {pages} {450} (\bibinfo {year} {1978})}\BibitemShut
  {NoStop}%
\bibitem [{\citenamefont {Zaslavsky}(1981)}]{Zaslavsky1981}%
  \BibitemOpen
  \bibfield  {author} {\bibinfo {author} {\bibfnamefont {G.~M.}\ \bibnamefont
  {Zaslavsky}},\ }\bibfield  {title} {\enquote {\bibinfo {title} {Stochasticity
  in quantum systems},}\ }\href@noop {} {\bibfield  {journal} {\bibinfo
  {journal} {Phys. Rep.}\ }\textbf {\bibinfo {volume} {80}},\ \bibinfo {pages}
  {157} (\bibinfo {year} {1981})}\BibitemShut {NoStop}%
\bibitem [{\citenamefont {Chirikov}\ \emph {et~al.}(1988)\citenamefont
  {Chirikov}, \citenamefont {Izrailev},\ and\ \citenamefont
  {Shepelyansky}}]{Chirikov1988}%
  \BibitemOpen
  \bibfield  {author} {\bibinfo {author} {\bibfnamefont {B.~V.}\ \bibnamefont
  {Chirikov}}, \bibinfo {author} {\bibfnamefont {F.~M.}\ \bibnamefont
  {Izrailev}}, \ and\ \bibinfo {author} {\bibfnamefont {D.~L.}\ \bibnamefont
  {Shepelyansky}},\ }\bibfield  {title} {\enquote {\bibinfo {title} {Quantum
  chaos: Localization vs. ergodicity},}\ }\href@noop {} {\bibfield  {journal}
  {\bibinfo  {journal} {Physica D}\ }\textbf {\bibinfo {volume} {33}},\
  \bibinfo {pages} {77--78} (\bibinfo {year} {1988})}\BibitemShut {NoStop}%
\bibitem [{\citenamefont {Izrailev}(1990)}]{Izrailev1990}%
  \BibitemOpen
  \bibfield  {author} {\bibinfo {author} {\bibfnamefont {F.~M.}\ \bibnamefont
  {Izrailev}},\ }\bibfield  {title} {\enquote {\bibinfo {title} {Simple models
  of quantum chaos: Spectrum and eigenfunctions},}\ }\href@noop {} {\bibfield
  {journal} {\bibinfo  {journal} {Phys. Rep.}\ }\textbf {\bibinfo {volume}
  {196}},\ \bibinfo {pages} {299--392} (\bibinfo {year} {1990})}\BibitemShut
  {NoStop}%
\bibitem [{\citenamefont {Fishman}\ \emph {et~al.}(1982)\citenamefont
  {Fishman}, \citenamefont {Grempel},\ and\ \citenamefont
  {Prange}}]{Fishman1982}%
  \BibitemOpen
  \bibfield  {author} {\bibinfo {author} {\bibfnamefont {S.}~\bibnamefont
  {Fishman}}, \bibinfo {author} {\bibfnamefont {D.~R.}\ \bibnamefont
  {Grempel}}, \ and\ \bibinfo {author} {\bibfnamefont {R.~E.}\ \bibnamefont
  {Prange}},\ }\bibfield  {title} {\enquote {\bibinfo {title} {Chaos, quantum
  recurrences, and anderson localization},}\ }\href@noop {} {\bibfield
  {journal} {\bibinfo  {journal} {Phys. Rev. Lett.}\ }\textbf {\bibinfo
  {volume} {49}},\ \bibinfo {pages} {509} (\bibinfo {year} {1982})}\BibitemShut
  {NoStop}%
\bibitem [{\citenamefont {Casati}\ \emph {et~al.}(1990)\citenamefont {Casati},
  \citenamefont {Molinari},\ and\ \citenamefont {Izrailev}}]{Casati1990}%
  \BibitemOpen
  \bibfield  {author} {\bibinfo {author} {\bibfnamefont {G.}~\bibnamefont
  {Casati}}, \bibinfo {author} {\bibfnamefont {L.}~\bibnamefont {Molinari}}, \
  and\ \bibinfo {author} {\bibfnamefont {F.~M.}\ \bibnamefont {Izrailev}},\
  }\bibfield  {title} {\enquote {\bibinfo {title} {Scaling properties of band
  random matrices},}\ }\href@noop {} {\bibfield  {journal} {\bibinfo  {journal}
  {Phys. Rev. Lett.}\ }\textbf {\bibinfo {volume} {64}},\ \bibinfo {pages}
  {1851--1854} (\bibinfo {year} {1990})}\BibitemShut {NoStop}%
\bibitem [{\citenamefont {Casati}\ \emph {et~al.}(1991)\citenamefont {Casati},
  \citenamefont {Izrailev},\ and\ \citenamefont {Molinari}}]{Casati1991}%
  \BibitemOpen
  \bibfield  {author} {\bibinfo {author} {\bibfnamefont {G.}~\bibnamefont
  {Casati}}, \bibinfo {author} {\bibfnamefont {F.~M.}\ \bibnamefont
  {Izrailev}}, \ and\ \bibinfo {author} {\bibfnamefont {L.}~\bibnamefont
  {Molinari}},\ }\bibfield  {title} {\enquote {\bibinfo {title} {Scaling
  properties of eigenvalue spacing distribution for band random matrices},}\
  }\href@noop {} {\bibfield  {journal} {\bibinfo  {journal} {J. Phys. A}\
  }\textbf {\bibinfo {volume} {24}},\ \bibinfo {pages} {4755} (\bibinfo {year}
  {1991})}\BibitemShut {NoStop}%
\bibitem [{\citenamefont {Zyczkowski}\ \emph {et~al.}(1992)\citenamefont
  {Zyczkowski}, \citenamefont {Lewenstein}, \citenamefont {Kus},\ and\
  \citenamefont {Izrailev}}]{Zyczkowski1992}%
  \BibitemOpen
  \bibfield  {author} {\bibinfo {author} {\bibfnamefont {K.}~\bibnamefont
  {Zyczkowski}}, \bibinfo {author} {\bibfnamefont {M.}~\bibnamefont
  {Lewenstein}}, \bibinfo {author} {\bibfnamefont {M.}~\bibnamefont {Kus}}, \
  and\ \bibinfo {author} {\bibfnamefont {F.~M.}\ \bibnamefont {Izrailev}},\
  }\bibfield  {title} {\enquote {\bibinfo {title} {Eigenvector statistics of
  random band matrices},}\ }\href@noop {} {\bibfield  {journal} {\bibinfo
  {journal} {Phys. Rev. A}\ }\textbf {\bibinfo {volume} {45}},\ \bibinfo
  {pages} {811--815} (\bibinfo {year} {1992})}\BibitemShut {NoStop}%
\bibitem [{\citenamefont {Izrailev}(1995)}]{Izrailev1995}%
  \BibitemOpen
  \bibfield  {author} {\bibinfo {author} {\bibfnamefont {F.~M.}\ \bibnamefont
  {Izrailev}},\ }\bibfield  {title} {\enquote {\bibinfo {title} {Quantum chaos,
  localization and band random matrices},}\ }in\ \href@noop {} {\emph {\bibinfo
  {booktitle} {Quantum Chaos: Between Order and Disorder}}},\ \bibinfo {editor}
  {edited by\ \bibinfo {editor} {\bibfnamefont {G.}~\bibnamefont {Casati}}\
  and\ \bibinfo {editor} {\bibfnamefont {B.}~\bibnamefont {Chirikov}}}\
  (\bibinfo  {publisher} {Cambridge Univ. Press},\ \bibinfo {year} {1995})\
  pp.\ \bibinfo {pages} {557--576}\BibitemShut {NoStop}%
\bibitem [{\citenamefont {Izrailev}\ \emph {et~al.}(1996)\citenamefont
  {Izrailev}, \citenamefont {Molinari},\ and\ \citenamefont
  {Zyczkovski}}]{Izrailev1996}%
  \BibitemOpen
  \bibfield  {author} {\bibinfo {author} {\bibfnamefont {F.~M.}\ \bibnamefont
  {Izrailev}}, \bibinfo {author} {\bibfnamefont {L.}~\bibnamefont {Molinari}},
  \ and\ \bibinfo {author} {\bibfnamefont {K.}~\bibnamefont {Zyczkovski}},\
  }\bibfield  {title} {\enquote {\bibinfo {title} {Periodic and non-periodic
  band random matrices: Structure of eigenstates},}\ }\href@noop {} {\bibfield
  {journal} {\bibinfo  {journal} {J. Phys. France}\ }\textbf {\bibinfo {volume}
  {6}},\ \bibinfo {pages} {455--468} (\bibinfo {year} {1996})}\BibitemShut
  {NoStop}%
\bibitem [{\citenamefont {Shepelyansky}(1986)}]{Shepelyansky1986}%
  \BibitemOpen
  \bibfield  {author} {\bibinfo {author} {\bibfnamefont {D.~L.}\ \bibnamefont
  {Shepelyansky}},\ }\bibfield  {title} {\enquote {\bibinfo {title}
  {Localization of quasienergy eigenfunctions in action space},}\ }\href@noop
  {} {\bibfield  {journal} {\bibinfo  {journal} {Phys. Rev. Lett.}\ }\textbf
  {\bibinfo {volume} {56}},\ \bibinfo {pages} {677--680} (\bibinfo {year}
  {1986})}\BibitemShut {NoStop}%
\bibitem [{\citenamefont {Casati}\ \emph {et~al.}(1993)\citenamefont {Casati},
  \citenamefont {Chirikov}, \citenamefont {Guarneri},\ and\ \citenamefont
  {Izrailev}}]{casati1993}%
  \BibitemOpen
  \bibfield  {author} {\bibinfo {author} {\bibfnamefont {G.}~\bibnamefont
  {Casati}}, \bibinfo {author} {\bibfnamefont {B.~V.}\ \bibnamefont
  {Chirikov}}, \bibinfo {author} {\bibfnamefont {I.}~\bibnamefont {Guarneri}},
  \ and\ \bibinfo {author} {\bibfnamefont {F.~M.}\ \bibnamefont {Izrailev}},\
  }\bibfield  {title} {\enquote {\bibinfo {title} {Band-random-matrix model for
  quantum localization in conservative systems},}\ }\href@noop {} {\bibfield
  {journal} {\bibinfo  {journal} {Phys. Rev. E}\ }\textbf {\bibinfo {volume}
  {48}},\ \bibinfo {pages} {R1613} (\bibinfo {year} {1993})}\BibitemShut
  {NoStop}%
\bibitem [{\citenamefont {Casati}\ \emph {et~al.}(1996)\citenamefont {Casati},
  \citenamefont {Chirikov}, \citenamefont {Guarneri},\ and\ \citenamefont
  {Izrailev}}]{casati1996}%
  \BibitemOpen
  \bibfield  {author} {\bibinfo {author} {\bibfnamefont {G.}~\bibnamefont
  {Casati}}, \bibinfo {author} {\bibfnamefont {B.~V.}\ \bibnamefont
  {Chirikov}}, \bibinfo {author} {\bibfnamefont {I.}~\bibnamefont {Guarneri}},
  \ and\ \bibinfo {author} {\bibfnamefont {F.~M.}\ \bibnamefont {Izrailev}},\
  }\bibfield  {title} {\enquote {\bibinfo {title} {Quantum ergodicity and
  localization in conservative systems: the wigner band random matrix model},}\
  }\href@noop {} {\bibfield  {journal} {\bibinfo  {journal} {Phys. Lett. A}\
  }\textbf {\bibinfo {volume} {223}},\ \bibinfo {pages} {430} (\bibinfo {year}
  {1996})}\BibitemShut {NoStop}%
\bibitem [{\citenamefont {Izrailev}(2001)}]{Izrailev2001}%
  \BibitemOpen
  \bibfield  {author} {\bibinfo {author} {\bibfnamefont {F.~M.}\ \bibnamefont
  {Izrailev}},\ }\bibfield  {title} {\enquote {\bibinfo {title}
  {Quantum-classical correspondence for isolated systems of interacting
  particles: Localization and ergodicity in energy space},}\ }\href@noop {}
  {\bibfield  {journal} {\bibinfo  {journal} {Phys. Scr.}\ }\textbf {\bibinfo
  {volume} {T90}},\ \bibinfo {pages} {95--104} (\bibinfo {year}
  {2001})}\BibitemShut {NoStop}%
\bibitem [{\citenamefont {Santos}\ \emph
  {et~al.}(2012{\natexlab{a}})\citenamefont {Santos}, \citenamefont
  {Borgonovi},\ and\ \citenamefont {Izrailev}}]{Santos2012PRL}%
  \BibitemOpen
  \bibfield  {author} {\bibinfo {author} {\bibfnamefont {L.~F.}\ \bibnamefont
  {Santos}}, \bibinfo {author} {\bibfnamefont {F.}~\bibnamefont {Borgonovi}}, \
  and\ \bibinfo {author} {\bibfnamefont {F.~M.}\ \bibnamefont {Izrailev}},\
  }\bibfield  {title} {\enquote {\bibinfo {title} {Chaos and statistical
  relaxation in quantum systems of interacting particles},}\ }\href@noop {}
  {\bibfield  {journal} {\bibinfo  {journal} {Phys. Rev. Lett.}\ }\textbf
  {\bibinfo {volume} {108}},\ \bibinfo {pages} {094102} (\bibinfo {year}
  {2012}{\natexlab{a}})}\BibitemShut {NoStop}%
\bibitem [{\citenamefont {Santos}\ \emph
  {et~al.}(2012{\natexlab{b}})\citenamefont {Santos}, \citenamefont
  {Borgonovi},\ and\ \citenamefont {Izrailev}}]{Santos2012PRE}%
  \BibitemOpen
  \bibfield  {author} {\bibinfo {author} {\bibfnamefont {L.~F.}\ \bibnamefont
  {Santos}}, \bibinfo {author} {\bibfnamefont {F.}~\bibnamefont {Borgonovi}}, \
  and\ \bibinfo {author} {\bibfnamefont {F.~M.}\ \bibnamefont {Izrailev}},\
  }\bibfield  {title} {\enquote {\bibinfo {title} {Onset of chaos and
  relaxation in isolated systems of interacting spins-1/2: energy shell
  approach},}\ }\href@noop {} {\bibfield  {journal} {\bibinfo  {journal} {Phys.
  Rev. E}\ }\textbf {\bibinfo {volume} {85}},\ \bibinfo {pages} {036209}
  (\bibinfo {year} {2012}{\natexlab{b}})}\BibitemShut {NoStop}%
\bibitem [{\citenamefont {Brody}\ \emph {et~al.}(1981)\citenamefont {Brody},
  \citenamefont {Flores}, \citenamefont {French}, \citenamefont {Mello},
  \citenamefont {Pandey},\ and\ \citenamefont {Wong}}]{Brody1981}%
  \BibitemOpen
  \bibfield  {author} {\bibinfo {author} {\bibfnamefont {T.~A.}\ \bibnamefont
  {Brody}}, \bibinfo {author} {\bibfnamefont {J.}~\bibnamefont {Flores}},
  \bibinfo {author} {\bibfnamefont {J.~B.}\ \bibnamefont {French}}, \bibinfo
  {author} {\bibfnamefont {P.~A.}\ \bibnamefont {Mello}}, \bibinfo {author}
  {\bibfnamefont {A.}~\bibnamefont {Pandey}}, \ and\ \bibinfo {author}
  {\bibfnamefont {S.~S.~M.}\ \bibnamefont {Wong}},\ }\bibfield  {title}
  {\enquote {\bibinfo {title} {Random-matrix physics -- spectrum and strength
  fluctuations},}\ }\href@noop {} {\bibfield  {journal} {\bibinfo  {journal}
  {Rev. Mod. Phys}\ }\textbf {\bibinfo {volume} {53}},\ \bibinfo {pages} {385}
  (\bibinfo {year} {1981})}\BibitemShut {NoStop}%
\bibitem [{\citenamefont {Kota}(2001)}]{Kota2001}%
  \BibitemOpen
  \bibfield  {author} {\bibinfo {author} {\bibfnamefont {V.~K.~B.}\
  \bibnamefont {Kota}},\ }\bibfield  {title} {\enquote {\bibinfo {title}
  {Embedded random matrix ensembles for complexity and chaos in finite
  interacting particle systems},}\ }\href@noop {} {\bibfield  {journal}
  {\bibinfo  {journal} {Phys. Rep.}\ }\textbf {\bibinfo {volume} {347}},\
  \bibinfo {pages} {223} (\bibinfo {year} {2001})}\BibitemShut {NoStop}%
\bibitem [{\citenamefont {Kota}(2014)}]{KotaBook}%
  \BibitemOpen
  \bibfield  {author} {\bibinfo {author} {\bibfnamefont {V.~K.~B.}\
  \bibnamefont {Kota}},\ }\href@noop {} {\emph {\bibinfo {title} {Lecture Notes
  in Physics, vol. 884}}}\ (\bibinfo  {publisher} {Springer},\ \bibinfo
  {address} {Heidelberg},\ \bibinfo {year} {2014})\BibitemShut {NoStop}%
\bibitem [{\citenamefont {Kitaev}({\natexlab{b}})}]{KitaevTALKkitp}%
  \BibitemOpen
  \bibfield  {author} {\bibinfo {author} {\bibfnamefont {A.}~\bibnamefont
  {Kitaev}},\ }\href@noop {} {\enquote {\bibinfo {title} {Kitp talk},}\ }
  ({\natexlab{b}}),\ \bibinfo {note}
  {http://online.kitp.ucsb.edu/online/entangled15/kitaev/}\BibitemShut
  {NoStop}%
\bibitem [{\citenamefont {Sachdev}\ and\ \citenamefont
  {Ye}(1993)}]{Sachdev1993}%
  \BibitemOpen
  \bibfield  {author} {\bibinfo {author} {\bibfnamefont {Subir}\ \bibnamefont
  {Sachdev}}\ and\ \bibinfo {author} {\bibfnamefont {Jinwu}\ \bibnamefont
  {Ye}},\ }\bibfield  {title} {\enquote {\bibinfo {title} {Gapless spin-fluid
  ground state in a random quantum heisenberg magnet},}\ }\href {\doibase
  10.1103/PhysRevLett.70.3339} {\bibfield  {journal} {\bibinfo  {journal}
  {Phys. Rev. Lett.}\ }\textbf {\bibinfo {volume} {70}},\ \bibinfo {pages}
  {3339--3342} (\bibinfo {year} {1993})}\BibitemShut {NoStop}%
\bibitem [{\citenamefont {Borgonovi}\ and\ \citenamefont
  {Izrailev}(2017)}]{Borgonoviaip}%
  \BibitemOpen
  \bibfield  {author} {\bibinfo {author} {\bibfnamefont {F.}~\bibnamefont
  {Borgonovi}}\ and\ \bibinfo {author} {\bibfnamefont {F.~M.}\ \bibnamefont
  {Izrailev}},\ }\bibfield  {title} {\enquote {\bibinfo {title} {Localized
  thermal states},}\ }in\ \href@noop {} {\emph {\bibinfo {booktitle}
  {Conference Proceedings AIP Publishing}}},\ \bibinfo {editor} {edited by\
  \bibinfo {editor} {\bibfnamefont {020003}\ \bibnamefont {1912}}}\ (\bibinfo
  {publisher} {AIP},\ \bibinfo {address} {New York},\ \bibinfo {year}
  {2017})\BibitemShut {NoStop}%
\bibitem [{\citenamefont {French}\ and\ \citenamefont
  {Wong}(1970)}]{French1970}%
  \BibitemOpen
  \bibfield  {author} {\bibinfo {author} {\bibfnamefont {J.~B.}\ \bibnamefont
  {French}}\ and\ \bibinfo {author} {\bibfnamefont {S.~S.~M.}\ \bibnamefont
  {Wong}},\ }\bibfield  {title} {\enquote {\bibinfo {title} {Validity of random
  matrix theories for many-particle systems},}\ }\href@noop {} {\bibfield
  {journal} {\bibinfo  {journal} {Phys. Lett. B}\ }\textbf {\bibinfo {volume}
  {33}},\ \bibinfo {pages} {449} (\bibinfo {year} {1970})}\BibitemShut
  {NoStop}%
\bibitem [{\citenamefont {Flambaum}\ and\ \citenamefont
  {Izrailev}(1997)}]{Flambaum1997}%
  \BibitemOpen
  \bibfield  {author} {\bibinfo {author} {\bibfnamefont {V.~V.}\ \bibnamefont
  {Flambaum}}\ and\ \bibinfo {author} {\bibfnamefont {F.~M.}\ \bibnamefont
  {Izrailev}},\ }\bibfield  {title} {\enquote {\bibinfo {title} {Statistical
  theory of finite {F}ermi systems based on the structure of chaotic
  eigenstates},}\ }\href@noop {} {\bibfield  {journal} {\bibinfo  {journal}
  {Phys. Rev. E}\ }\textbf {\bibinfo {volume} {56}},\ \bibinfo {pages} {5144}
  (\bibinfo {year} {1997})}\BibitemShut {NoStop}%
\bibitem [{\citenamefont {Altshuler}\ \emph {et~al.}(1997)\citenamefont
  {Altshuler}, \citenamefont {Gefen}, \citenamefont {Kamenev},\ and\
  \citenamefont {Levitov}}]{Altshuler1997}%
  \BibitemOpen
  \bibfield  {author} {\bibinfo {author} {\bibfnamefont {Boris~L.}\
  \bibnamefont {Altshuler}}, \bibinfo {author} {\bibfnamefont {Yuval}\
  \bibnamefont {Gefen}}, \bibinfo {author} {\bibfnamefont {Alex}\ \bibnamefont
  {Kamenev}}, \ and\ \bibinfo {author} {\bibfnamefont {Leonid~S.}\ \bibnamefont
  {Levitov}},\ }\bibfield  {title} {\enquote {\bibinfo {title} {Quasiparticle
  lifetime in a finite system: A nonperturbative approach},}\ }\href {\doibase
  10.1103/PhysRevLett.78.2803} {\bibfield  {journal} {\bibinfo  {journal}
  {Phys. Rev. Lett.}\ }\textbf {\bibinfo {volume} {78}},\ \bibinfo {pages}
  {2803--2806} (\bibinfo {year} {1997})}\BibitemShut {NoStop}%
\bibitem [{\citenamefont {Kota}\ and\ \citenamefont
  {Sahu}(2001)}]{Kota2001PRE}%
  \BibitemOpen
  \bibfield  {author} {\bibinfo {author} {\bibfnamefont {V.~K.~B.}\
  \bibnamefont {Kota}}\ and\ \bibinfo {author} {\bibfnamefont {R.}~\bibnamefont
  {Sahu}},\ }\bibfield  {title} {\enquote {\bibinfo {title} {Structure of wave
  functions in (1+2)-body random matrix ensembles},}\ }\href {\doibase
  10.1103/PhysRevE.64.016219} {\bibfield  {journal} {\bibinfo  {journal} {Phys.
  Rev. E}\ }\textbf {\bibinfo {volume} {64}},\ \bibinfo {pages} {016219}
  (\bibinfo {year} {2001})}\BibitemShut {NoStop}%
\bibitem [{\citenamefont {Benet}\ and\ \citenamefont
  {Weidenm\"uller}(2003)}]{Benet2003}%
  \BibitemOpen
  \bibfield  {author} {\bibinfo {author} {\bibfnamefont {L}~\bibnamefont
  {Benet}}\ and\ \bibinfo {author} {\bibfnamefont {H~A}\ \bibnamefont
  {Weidenm\"uller}},\ }\bibfield  {title} {\enquote {\bibinfo {title} {Review
  of the k -body embedded ensembles of gaussian random matrices},}\ }\href
  {http://stacks.iop.org/0305-4470/36/i=12/a=340} {\bibfield  {journal}
  {\bibinfo  {journal} {Journal of Physics A: Mathematical and General}\
  }\textbf {\bibinfo {volume} {36}},\ \bibinfo {pages} {3569} (\bibinfo {year}
  {2003})}\BibitemShut {NoStop}%
\bibitem [{\citenamefont {Santos}(2009)}]{Santos2009JMP}%
  \BibitemOpen
  \bibfield  {author} {\bibinfo {author} {\bibfnamefont {L.~F.}\ \bibnamefont
  {Santos}},\ }\bibfield  {title} {\enquote {\bibinfo {title} {Transport and
  control in one-dimensional systems},}\ }\href@noop {} {\bibfield  {journal}
  {\bibinfo  {journal} {J. Math. Phys}\ }\textbf {\bibinfo {volume} {50}},\
  \bibinfo {pages} {095211} (\bibinfo {year} {2009})}\BibitemShut {NoStop}%
\bibitem [{\citenamefont {Torres-Herrera}\ \emph {et~al.}(2015)\citenamefont
  {Torres-Herrera}, \citenamefont {Kollmar},\ and\ \citenamefont
  {Santos}}]{TorresKollmar2015}%
  \BibitemOpen
  \bibfield  {author} {\bibinfo {author} {\bibfnamefont {E.~J.}\ \bibnamefont
  {Torres-Herrera}}, \bibinfo {author} {\bibfnamefont {D.}~\bibnamefont
  {Kollmar}}, \ and\ \bibinfo {author} {\bibfnamefont {Lea~F.}\ \bibnamefont
  {Santos}},\ }\bibfield  {title} {\enquote {\bibinfo {title} {Relaxation and
  thermalization of isolated many-body quantum systems},}\ }\href@noop {}
  {\bibfield  {journal} {\bibinfo  {journal} {Phys. Scr. T}\ }\textbf {\bibinfo
  {volume} {165}},\ \bibinfo {pages} {014018} (\bibinfo {year}
  {2015})}\BibitemShut {NoStop}%
\bibitem [{\citenamefont {Flambaum}\ and\ \citenamefont
  {Izrailev}(2001)}]{Flambaum2001b}%
  \BibitemOpen
  \bibfield  {author} {\bibinfo {author} {\bibfnamefont {V.~V.}\ \bibnamefont
  {Flambaum}}\ and\ \bibinfo {author} {\bibfnamefont {F.~M.}\ \bibnamefont
  {Izrailev}},\ }\bibfield  {title} {\enquote {\bibinfo {title} {Entropy
  production and wave packet dynamics in the fock space of closed chaotic
  many-body systems},}\ }\href@noop {} {\bibfield  {journal} {\bibinfo
  {journal} {Phys. Rev. E}\ }\textbf {\bibinfo {volume} {64}},\ \bibinfo
  {pages} {036220} (\bibinfo {year} {2001})}\BibitemShut {NoStop}%
\bibitem [{\citenamefont {Borgonovi}\ and\ \citenamefont
  {Izrailev}(2019)}]{Borgonovi2019b}%
  \BibitemOpen
  \bibfield  {author} {\bibinfo {author} {\bibfnamefont {Fausto}\ \bibnamefont
  {Borgonovi}}\ and\ \bibinfo {author} {\bibnamefont {Izrailev}},\ }\bibfield
  {title} {\enquote {\bibinfo {title} {Emergence of correlations in the process
  of thermalization of interacting bosons},}\ }\href {\doibase
  10.1103/PhysRevE.99.012115} {\bibfield  {journal} {\bibinfo  {journal} {Phys.
  Rev. E}\ }\textbf {\bibinfo {volume} {99}},\ \bibinfo {pages} {012115}
  (\bibinfo {year} {2019})}\BibitemShut {NoStop}%
\bibitem [{\citenamefont {Kukuljan}\ \emph {et~al.}(2017)\citenamefont
  {Kukuljan}, \citenamefont {Grozdanov},\ and\ \citenamefont
  {Prosen}}]{Kukuljan2017}%
  \BibitemOpen
  \bibfield  {author} {\bibinfo {author} {\bibfnamefont {I.}~\bibnamefont
  {Kukuljan}}, \bibinfo {author} {\bibfnamefont {S.}~\bibnamefont {Grozdanov}},
  \ and\ \bibinfo {author} {\bibfnamefont {T.}~\bibnamefont {Prosen}},\
  }\bibfield  {title} {\enquote {\bibinfo {title} {Weak quantum chaos},}\
  }\href {\doibase 10.1103/PhysRevB.96.060301} {\bibfield  {journal} {\bibinfo
  {journal} {Phys. Rev. B}\ }\textbf {\bibinfo {volume} {96}},\ \bibinfo
  {pages} {060301} (\bibinfo {year} {2017})}\BibitemShut {NoStop}%
\bibitem [{\citenamefont {Garcia-Mata}\ \emph {et~al.}()\citenamefont
  {Garcia-Mata}, \citenamefont {Saraceno}, \citenamefont {Jalabert},
  \citenamefont {Roncaglia},\ and\ \citenamefont {Wisniacki}}]{MataARXIV}%
  \BibitemOpen
  \bibfield  {author} {\bibinfo {author} {\bibfnamefont {Ignacio}\ \bibnamefont
  {Garcia-Mata}}, \bibinfo {author} {\bibfnamefont {Marcos}\ \bibnamefont
  {Saraceno}}, \bibinfo {author} {\bibfnamefont {Rodolfo~A.}\ \bibnamefont
  {Jalabert}}, \bibinfo {author} {\bibfnamefont {Augusto~J.}\ \bibnamefont
  {Roncaglia}}, \ and\ \bibinfo {author} {\bibfnamefont {Diego~A.}\
  \bibnamefont {Wisniacki}},\ }\href@noop {} {\enquote {\bibinfo {title} {Chaos
  signatures in the short and long time behavior of the out-of-time ordered
  correlator},}\ }\bibinfo {note} {ArXiv:1806.04281}\BibitemShut {NoStop}%
\bibitem [{\citenamefont {Torres-Herrera}\ and\ \citenamefont
  {Santos}(2017{\natexlab{a}})}]{Torres2017}%
  \BibitemOpen
  \bibfield  {author} {\bibinfo {author} {\bibfnamefont {E.~J.}\ \bibnamefont
  {Torres-Herrera}}\ and\ \bibinfo {author} {\bibfnamefont {L.~F.}\
  \bibnamefont {Santos}},\ }\bibfield  {title} {\enquote {\bibinfo {title}
  {Extended nonergodic states in disordered many-body quantum systems},}\
  }\href {\doibase 10.1002/andp.201600284} {\bibfield  {journal} {\bibinfo
  {journal} {Ann. Phys. (Berlin)}\ }\textbf {\bibinfo {volume} {529}},\
  \bibinfo {pages} {1600284} (\bibinfo {year}
  {2017}{\natexlab{a}})}\BibitemShut {NoStop}%
\bibitem [{\citenamefont {Torres-Herrera}\ and\ \citenamefont
  {Santos}(2017{\natexlab{b}})}]{Torres2017PTR}%
  \BibitemOpen
  \bibfield  {author} {\bibinfo {author} {\bibfnamefont {E.~J.}\ \bibnamefont
  {Torres-Herrera}}\ and\ \bibinfo {author} {\bibfnamefont {L.~F.}\
  \bibnamefont {Santos}},\ }\bibfield  {title} {\enquote {\bibinfo {title}
  {Dynamical manifestations of quantum chaos: Correlation hole and bulge},}\
  }\href@noop {} {\bibfield  {journal} {\bibinfo  {journal} {Phil. Trans. R.
  Soc. A}\ }\textbf {\bibinfo {volume} {375}},\ \bibinfo {pages} {20160434}
  (\bibinfo {year} {2017}{\natexlab{b}})}\BibitemShut {NoStop}%
\bibitem [{\citenamefont {Torres-Herrera}\ \emph {et~al.}(2018)\citenamefont
  {Torres-Herrera}, \citenamefont {Garc\'{\i}a-Garc\'{\i}a},\ and\
  \citenamefont {Santos}}]{Torres2018}%
  \BibitemOpen
  \bibfield  {author} {\bibinfo {author} {\bibfnamefont {E.~J.}\ \bibnamefont
  {Torres-Herrera}}, \bibinfo {author} {\bibfnamefont {Antonio~M.}\
  \bibnamefont {Garc\'{\i}a-Garc\'{\i}a}}, \ and\ \bibinfo {author}
  {\bibfnamefont {Lea~F.}\ \bibnamefont {Santos}},\ }\bibfield  {title}
  {\enquote {\bibinfo {title} {Generic dynamical features of quenched
  interacting quantum systems: Survival probability, density imbalance, and
  out-of-time-ordered correlator},}\ }\href {\doibase
  10.1103/PhysRevB.97.060303} {\bibfield  {journal} {\bibinfo  {journal} {Phys.
  Rev. B}\ }\textbf {\bibinfo {volume} {97}},\ \bibinfo {pages} {060303}
  (\bibinfo {year} {2018})}\BibitemShut {NoStop}%
\bibitem [{\citenamefont {F.~Borgonovi}\ and\ \citenamefont
  {Izrailev}(1998)}]{Borgonovi1998b}%
  \BibitemOpen
  \bibfield  {author} {\bibinfo {author} {\bibfnamefont {I.~Guarneri}\
  \bibnamefont {F.~Borgonovi}}\ and\ \bibinfo {author} {\bibfnamefont {F.~M.}\
  \bibnamefont {Izrailev}},\ }\bibfield  {title} {\enquote {\bibinfo {title}
  {Quantum-classical correspondence in energy space: Two interacting
  spin-particles},}\ }\href@noop {} {\bibfield  {journal} {\bibinfo  {journal}
  {Phys. Rev. E}\ }\textbf {\bibinfo {volume} {57}},\ \bibinfo {pages}
  {5291--5302} (\bibinfo {year} {1998})}\BibitemShut {NoStop}%
\bibitem [{\citenamefont {G.A.Luna-Acosta}\ \emph {et~al.}(2000)\citenamefont
  {G.A.Luna-Acosta}, \citenamefont {J.A.M\'{e}ndes-Berm\'{u}dez},\ and\
  \citenamefont {F.M.Izrailev}}]{Luna2000}%
  \BibitemOpen
  \bibfield  {author} {\bibinfo {author} {\bibnamefont {G.A.Luna-Acosta}},
  \bibinfo {author} {\bibnamefont {J.A.M\'{e}ndes-Berm\'{u}dez}}, \ and\
  \bibinfo {author} {\bibnamefont {F.M.Izrailev}},\ }\bibfield  {title}
  {\enquote {\bibinfo {title} {Quantum-classical correspondence for local
  density of states and eigenfuctions of a chaotic periodic billiard},}\
  }\href@noop {} {\bibfield  {journal} {\bibinfo  {journal} {Phys. Lett. A}\
  }\textbf {\bibinfo {volume} {274}},\ \bibinfo {pages} {192--199} (\bibinfo
  {year} {2000})}\BibitemShut {NoStop}%
\bibitem [{\citenamefont {Luna-Acosta}\ \emph {et~al.}(2001)\citenamefont
  {Luna-Acosta}, \citenamefont {M\'{e}ndes-Berm\'{u}dez},\ and\ \citenamefont
  {Izrailev}}]{Luna2001}%
  \BibitemOpen
  \bibfield  {author} {\bibinfo {author} {\bibfnamefont {G.~A.}\ \bibnamefont
  {Luna-Acosta}}, \bibinfo {author} {\bibfnamefont {J.~A.}\ \bibnamefont
  {M\'{e}ndes-Berm\'{u}dez}}, \ and\ \bibinfo {author} {\bibfnamefont {F.~M.}\
  \bibnamefont {Izrailev}},\ }\bibfield  {title} {\enquote {\bibinfo {title}
  {Periodic chaotic billiards: Quantum-classical correspondence in energy
  space},}\ }\href@noop {} {\bibfield  {journal} {\bibinfo  {journal} {Phys.
  Rev. E}\ }\textbf {\bibinfo {volume} {64}},\ \bibinfo {pages} {036206}
  (\bibinfo {year} {2001})}\BibitemShut {NoStop}%
\bibitem [{\citenamefont {Luna-Acosta}\ \emph {et~al.}(2002)\citenamefont
  {Luna-Acosta}, \citenamefont {M\'{e}ndes-Berm\'{u}dez},\ and\ \citenamefont
  {Izrailev}}]{Luna2002}%
  \BibitemOpen
  \bibfield  {author} {\bibinfo {author} {\bibfnamefont {G.~A.}\ \bibnamefont
  {Luna-Acosta}}, \bibinfo {author} {\bibfnamefont {J.~A.}\ \bibnamefont
  {M\'{e}ndes-Berm\'{u}dez}}, \ and\ \bibinfo {author} {\bibfnamefont {F.~M.}\
  \bibnamefont {Izrailev}},\ }\bibfield  {title} {\enquote {\bibinfo {title}
  {Chaotic electron motion in superlattices. quantum-classical correspondence
  of the structure of eigenstates and {LDOS}},}\ }\href@noop {} {\bibfield
  {journal} {\bibinfo  {journal} {Physica E}\ }\textbf {\bibinfo {volume}
  {12}},\ \bibinfo {pages} {267} (\bibinfo {year} {2002})}\BibitemShut
  {NoStop}%
\end{thebibliography}%

\end{document}